# Yttrium Superhydrides Revisited:

# Advanced Experimental and Theoretical Studies of $YH_6$, $YH_9$ and $YH_{10}$

Dmitrii V. Semenok,[1, †, *] Pedro N. Ferreira,[2, †, *] Di Zhou,[1, †] Fabian Jőbstl,[3] Andrey V. Sadakov,[4] Kirill S. Pervakov,[4] Burkhan I. Massalimov,[4] Toni Helm,[5] Ryosuke Akashi [6], Vladimir M. Pudalov,[4,9] Viktor V. Struzhkin,[7, 8] Christoph Heil,[2] and Ivan A. Troyan[7, *]

[1] Center for High Pressure Science & Technology Advanced Research, Bldg. 8E, ZPark, 10 Xibeiwang East Rd, Haidian District, Beijing, 100193, China

[2] Institute of Theoretical and Computational Physics, Graz University of Technology, NAWI Graz, 8010, Graz, Austria

[3] University of Vienna, Faculty of Physics, Kolingasse 14-16, Vienna, Austria

[4] V. L. Ginzburg Center for High-Temperature Superconductivity and Quantum Materials, 53 Leninsky Prospekt, building 10, Moscow 119991, Russia

[5] Hochfeld-Magnetlabor Dresden (HLD-EMFL) and Würzburg-Dresden Cluster of Excellence, Helmholtz-Zentrum Dresden-Rossendorf (HZDR), Bautzner Landstrase 400, Dresden 01328, Germany

[6] Department of Applied Physics and Physico-Informatics, Keio University, 3-14-1 Hiyoshi, Yokohama, Kanagawa 223-0061, Japan

[7] Shanghai Key Laboratory of Material Frontiers Research in Extreme Environments (MFree), Shanghai Advanced Research in Physical Sciences (SHARPS), 68 Huatuo Rd, Bldg 3 Pudong, Shanghai 201203, China

[8] Center for High Pressure Science & Technology Advanced Research, 1690 Cailun Rd, Bldg 6, Pudong, Shanghai 201203, China

[9] National Research University Higher School of Economics, Moscow 101000, Russia

*Corresponding authors: Dmitrii V. Semenok (dmitrii.semenok@hpstar.ac.cn), Pedro N. Ferreira (nunesferreira@tugraz.at), and Ivan A. Troyan (itrojan@sharps.ac.cn)

† These authors contributed equally

## Abstract

Yttrium polyhydrides are benchmark materials in high-pressure superconductivity, yet several key properties of the Y-H system remain insufficiently characterized. Here we combine contact transport, contactless radio-frequency measurements, pulsed-field experiments, and first-principles calculations to reinvestigate $YH_6$, $YH_9$, and $YH_{10}$ in the pressure range 140-213 GPa. Yttrium hydrides $YH_6$ ($T_c$ = 218–226 K) and $YH_9$ ($T_c$ = 235–243 K) demonstrate narrow superconducting transitions ($\Delta T_c \approx$ 2–5 K), approaching the limit imposed by thermal fluctuations. Pulsed-field measurements on $YH_6$ up to 60 T establish an extended superconducting phase diagram with a linear slope $dB_{c2}/dT = -0.52$ T/K, pronounced transition broadening above 30 T, and negligible normal-state magnetoresistance. We report the radio-frequency AC susceptibility study of $YH_6$, providing evidence for superconductivity via high-frequency field screening in a contactless geometry. Experiments involving Pd incorporation, Pd thin-film sputtering, and Al alloying show strong suppression of high-temperature superconductivity, with no transitions detected above 78-120 K. Finally, using density-functional theory with the stochastic self-consistent harmonic approximation, superconducting density-functional theory, and full-bandwidth Migdal-Eliashberg calculations, we show that anharmonic effects substantially reduce the predicted $T_c$ of cubic $YH_{10}$ to approximately 260–270 K. These results strongly disfavor room-temperature superconductivity in binary yttrium superhydrides.

## Introduction

The yttrium–hydrogen system ranks second only to lanthanum–hydrogen in its importance for superhydride research. Within this system, $Fm\bar{3}m$-$YH_{10}$ has been theoretically predicted to exhibit room-temperature superconductivity up to 326 K [1,2], while the thermodynamically stable $P6_3/mmc$-$YH_9$ phase, which retains stability across an exceptionally wide range of pressures and temperatures, has been experimentally confirmed to superconduct with a maximum critical temperature of 243 K [3]. The next member of the yttrium superhydride family is $Im\bar{3}m$-$YH_6$ (max $T_c$ = 224–226 K [3-6]), which is frequently co-synthesized with the *cF*80 phase $Y_4H_{25\pm x}$ [7], a compound with an unsolved hydrogen sublattice that is very often encountered among lanthanide polyhydrides. The latter likely accounts for the secondary steps commonly observed in the resistive superconducting transitions of $YH_x$ samples, and is expected to superconduct above 200 K. Further members of the series include superconducting $YH_4$ (max $T_c$ = 80–88 K [5,6,8]) and A15-type $Pm\bar{3}n$-$Y_4H_{23}$ [7,9] (synthesized at 125-138 GPa), for which a critical temperature of 90–100 K is anticipated based on the superconducting properties of the similar $La_4H_{23}$ [10,11]. Sometimes, the formation of different yttrium hydrides is observed, for example, $YH_7$ [4-6] (max $T_c$ = 29 K ), $Y_3H_{11}$, $YH_3$, $Y_2H_9$, and $Y_{13}H_{75}$ [7], however, their superconducting properties are weakly expressed or unknown.

Among all stable binary metal hydrides studied to date, the cubic $XH_{10}$ compositions — of which $LaH_{10}$ [12] is the archetype — display the most favorable superconducting properties. However, a recurring pattern across multiple metal–hydrogen systems is that the hexagonal $XH_9$ stoichiometry, rather than the cubic $XH_{10}$, that prevails as the thermodynamically stable phase at pressures of ≈1–2 Mbar. In the Y-H system itself, the computationally predicted the room-temperature superconductor $YH_{10}$ has never been experimentally realized to date. Instead, hexagonal $YH_9$ is the stable phase across the broad pressure window of 200–400 GPa [3], with a possible further distortion to a lower-symmetry $P1$-$Y_4H_{36}$ superstructure [13]. Analogous behavior is observed in the thorium–hydrogen system: below approximately 94 GPa, the cubic $ThH_{10\pm x}$ phase distorts, progressively loses hydrogen, and decomposes into the more stable $P6_3/mmc$-$ThH_9$ — a phase that routinely appears as a persistent impurity in thorium decahydride syntheses [14]. A closely parallel situation exists in the cerium–hydrogen system, where the $CeH_9$–$CeH_{10}$ phase pair has been characterized at relatively accessible pressures of 70–120 GPa [15]. There, hexagonal $CeH_9$ is by far the most readily obtained product, exhibiting a monoclinic $C2/c$ distortion and retaining thermodynamic stability and superconductivity down to approximately 70 GPa.

**Table 1.** Experimentally determined parameters of the superconducting state for yttrium superhydrides. $\theta_D$, $\lambda_{BG}$ denote the parameters of the electron-phonon interaction, determined by fitting the *R(T)* data to formula (1) and then applying the Allen-Dynes formula with $\mu^*$ = 0.1.

| Compound | $P6_3/mmc$-$YH_9$ | $Im\bar{3}m$-$YH_6$ | $I4/mmm$-$YH_4$ |
|---|---|---|---|
| Stability region, GPa | 200-410 | 160-240 | 155-180 |
| $\lambda_{BG}$ | 2.66 [16] | 2.3-2.7 | 1.3* |
| $\theta_D$, K | 1160-1275 | 1200-1345 | ~910* |
| max $T_c$, K | 243 | 224-226 | 82-88 |
| max $B_{c2}(0)$ | 100-120 | 115-157 | 14-22 |

* Best fit is at $p$ = 2 in eq. (1), data is taken from Ref. [5]

In this work, we carried out an experimental and theoretical investigation of several previously unexplored properties of yttrium hydrides, including studies of the magnetic phase diagram of $YH_6$ in pulsed magnetic fields up to 60 T. We also investigated for the first time the diamagnetic response of yttrium hexahydride $YH_6$, synthesized along a non-standard route from $YH_2$/Y(AB)$_3$ at 165-180 GPa, using the radio-frequency (RF) transmission technique [17-20] in the range of 555 kHz - 5 MHz. The change in the transmitted RF signal below $T_c$ is consistent with the expulsion of the AC magnetic field from the sample.[21] We also reproduced the synthesis of $YH_9$ and confirmed its decomposition into $YH_6$ below 2 Mbar. Most importantly, using ab initio calculations that include quantum anharmonic lattice effects via the stochastic self-consistent harmonic approximation (SSCHA), combined with full-bandwidth Migdal–Eliashberg (ME) theory, we show that the critical temperature of $YH_{10}$ was significantly overestimated in earlier theoretical studies. Anharmonicity suppresses the soft hydrogen phonon modes and reduces the electron–phonon coupling, lowering the predicted $T_c$ from ~310–326 K to 260–270 K at 250 GPa — only slightly above $T_c(LaH_{10})$ and below room temperature.

*I. Experimental synthesis and electrical-transport properties of $YH_6$ and $YH_9$*

We investigated samples of yttrium hydrides in DAC Y1 (pulsed-field DAC, NiCrAl, $d$ = 15 mm), and DAC Y2 (steady-field BeCu, $d$ = 25 mm), both with a diamond-anvil culet diameter of 50 μm; the sample size was about 30-35 μm, and the thickness of ≈1-2 μm. The yttrium sample, pre-compressed to the thickness 1-2 μm, was loaded into the center of the culet together with $NH_3BH_3$, which served as the hydrogen source. We applied a nonmagnetic composite $CaF_2$/epoxy/tungsten insulating gasket to prevent electrical shorting between the electrodes via gasket material. Double-layer Ta/Au electrodes were brought into contact with the sample before closing the DAC and applying pressures up to 195-213 GPa. Laser heating was performed with an Nd:YAG fiber laser (1064 nm) with 200-300 ms pulses.

A study of the DAC Y1 revealed that the sample exhibits a very sharp superconducting transition at 218 K and 195 GPa, with a width of only 2 K (Figure 1a), and near-zero residual resistance (Supporting Figures S3-S4). The non-superconducting part of $R(T)$ allows us to estimate the Debye temperature using a Bloch–Grüneisen fit as $\theta_D$ = 1345 K. Using the Allen-Dynes formula[22], we estimate $\lambda_{BG} \approx 2.3$ ($\mu^* = 0.1$, Figure 1a). Metal superhydrides with an atomic hydrogen sublattice are normal metals whose normal-state electrical resistance is governed by electron scattering on phonons and defects. Their transport behavior can be described by the Bloch–Grüneisen formula [23,24]

$$R(T) = R_0 + A\left(\frac{T}{\theta_D}\right)^p \int_0^{\frac{\theta_D}{T}} \frac{x^p}{(e^x-1)(1-e^{-x})}\,dx, \qquad (1)$$

where the metal phase ratio $A$, the Debye temperature $\theta_D$, exponent $p$ and the residual resistance $R_0$ can be found using the least square method. For the electron-phonon scattering, $p = 5$. In the ideal case (e.g., Figures 1a, d), the normal-state resistance $R(T)$ is well described by this formula [25,26]. However, phonon scattering is not the only mechanism. Hydrides are nanodispersed materials with grain sizes of about 30–50 nm[27,28]. These materials exhibit fine powder X-ray diffraction even from samples with thicknesses of ≈1–2 μm and an X-ray beam of the compatible diameter. As a result, the contribution of grain boundaries can be significant, as is the case, for example, in Pd- and Al-doped yttrium hydrides (Figures 3c, g). This also explains why, in many cases, $R(T)$ of hydrides cannot be described by Eq. (1) even with variable $p \neq 5$ [29].

In pulsed magnetic fields above 20–25 T, the superconducting transition in $Im\bar{3}m$-$YH_6$ broadens significantly; at the same time the sample exhibits near-zero averaged magnetoresistance (Figure 1b). The temperature dependence of the upper critical magnetic field $B_{c2}(T)$ is linear with $dB_{c2}/dT$ = –0.52 T/K, regardless of the criterion used ($R_{50\%}$ - see Figure 1b, $R_{90\%}$ - see Figure 1c). Given the good agreement between the obtained $T_c$ and previously published data for $YH_6$, and considering the high asymmetry and very low opening angle of the pulsed field DACs, the X-ray diffraction experiments were not performed on this sample.

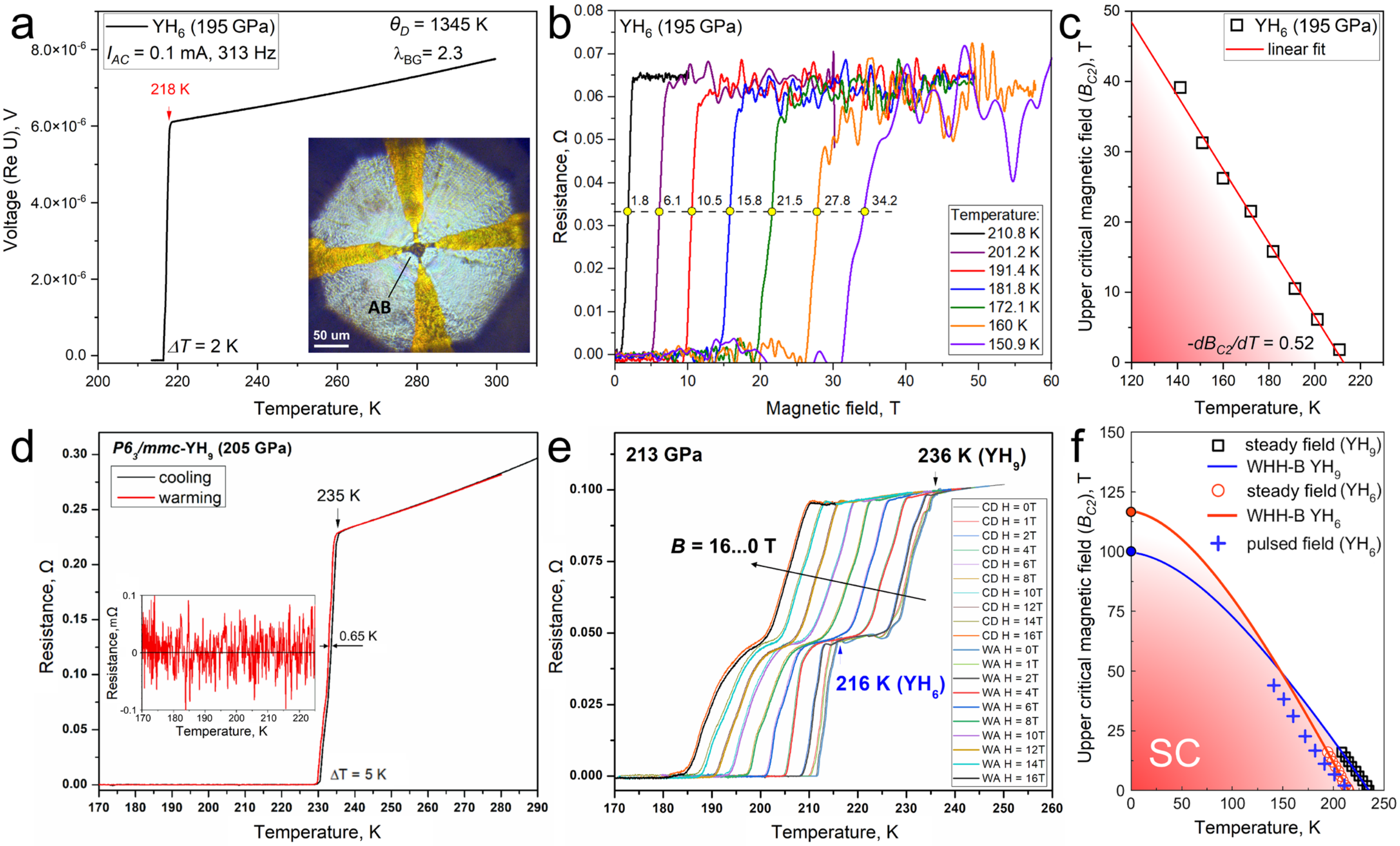

**Figure 1.** Electric transport measurements of yttrium hydrides $YH_6$ (DAC Y1) and $YH_9$ (DAC Y2) in steady and pulsed magnetic fields up to 16 and 60 Tesla, respectively. (a) Temperature dependence of the electric voltage drop (real component, Re U) on the $YH_6$ sample at 195 GPa, measured in the AC mode at a frequency of 313 Hz and a current $I_{AC}$ = 0.1 mA. Warming cycle. Inset: photograph of the DAC Y1 high-pressure chamber, which shows the sample after laser heating and Ta/Au electrodes. (b) Dependence of the resistance (real component, Re $Z$) of the DAC Y1 sample on magnetic field for a series of fixed temperatures from 151 K to 211 K, measured at a frequency of 7.77 kHz and a current $I_{AC}$ = 2 mA. (c) Temperature dependence of the upper critical field $B_{c2}(T)$ for the $YH_6$ sample in DAC Y1 obtained from pulsed measurements. (d) Temperature dependence of the electrical resistance of the $YH_9$ sample (DAC Y2) at 205 GPa measured in the DC (delta mode) mode at $I_{DC}$ = 0.1 mA. The warming cycle is shown by the red curve, the cooling cycle is shown by the black curve. Inset: near-zero resistance below 230 K, which is a noise track. (e) Temperature dependence of the electrical resistance of the DAC Y2 sample (mixture of $YH_6$ and $YH_9$) at 213 GPa after additional laser heating measured in the DC delta mode at $I_{DC}$ = 0.1 mA in magnetic fields from 0 to 16 Tesla in warming and cooling cycles. (f) Magnetic phase diagram of $YH_6$ and $YH_9$, with experimental data points obtained in pulsed and static magnetic fields and their extrapolation using the simplified Werthamer-Helfand-Hohenberg-Baumgartner (WHH-B) formula.

A distinctive feature of the DAC Y1 sample is the narrow superconducting transition in $YH_6$ at 218 K (195 GPa). The width of a superconducting transition in hydrides cannot be arbitrarily narrow: it is fundamentally limited by strong thermal fluctuations, inhomogeneities, vortex pinning, and quantum fluctuations. The influence of thermal fluctuations ($th$) in a magnetic field $B$ can be derived from the Bardeen–Cooper–Schrieffer–Migdal–Eliashberg (BCS-ME) theory: $\Delta T_c{}^{th}(B) = aB^{\frac{2}{3}} + \Delta T_c{}^{th}(0)$ [30]. In the zero-field limit, it is quantified by the Ginzburg–Levanyuk number [31], which we previously determined for $YH_6$ [32]

$$Gi = \frac{\Delta T_c{}^{th}}{T_c} \approx (3-7)\cdot 10^{-3}. \qquad (2)$$

This leads to a minimum transition width $\Delta T_c{}^{th} = 0.6 - 1.5$ K, which is reached with good accuracy in our experiment with DAC Y1 (Figure 1a). In general, $Gi \propto \kappa^4$, where $\kappa$ – is the Ginzburg-Landau parameter, in hydrides takes values of $10^{-2} - 10^{-3}$, [32,33] which fundamentally limits the transition width in the best hydrides to a value of 0.3–3 K. It also limits the temperature resolution of a hypothetical hydride-based superconducting thermometer to $\Delta T = 3-30$ μK, assuming a minimum measurable change in electrical resistance of about 10 μΩ.

A second sample (DAC Y2) was prepared similarly using a BeCu DAC (outer diameter $d$ = 25 mm). After laser heating at 205 GPa, the sample exhibited a narrow (5 K) superconducting transition with a minimal thermal hysteresis (0.65 K) below $T_c$(onset) = 235 K (Figure 1d). This is slightly lower than the maximum of 243 K achieved in an earlier work near the $YH_9$ stability boundary around 200 GPa [3]. Subsequently, we increased the pressure in DAC Y2 to 213 GPa, and sample was reheated by IR laser; this did not significantly change $T_c(YH_9)$ ≈ 236 K, but led to the appearance of an additional step at 216 K corresponding to $YH_6$, thereby indicating the two-phase composition of the DAC Y2 sample (Figure 1e). The transport behavior of this sample in magnetic fields (Figure 1e) shows that the extrapolated $B_{c2}(0)$ of $YH_9$ does not exceed 100 Tesla, which seems to be lower than the $B_{c2}$ of $YH_6$ (Figure 1f).

In the absence of experimental data at 4.2 K and below for $YH_6$ and $YH_9$, the upper critical field $B_{c2}(0)$ can be estimated using models based on the Migdal-Eliashberg (ME) theory. Carbotte [34] proposed formulas for $B_{c2}(0)$ in an ideal crystal ("clean" limit). In the “clean” limit, Ref. [34], eq. 7.17 reads

$$B_{c2}^{cl}(0) = \left(\frac{\pi}{2}\right)^2 e^{-\gamma+2}\,\frac{2T_c^2(1+\lambda)^2}{eV_F^2}\left(1 + 1.44\frac{T_c}{\omega_{log}}\right), \quad (3)$$

where γ = 0.577, e – is the absolute electron charge, $V_F$ – the averaged Fermi velocity. In the dirty limit, Ref. [34], eq. 7.11 reads

$$B_{c2}^{di}(0) = \frac{\pi}{2}e^{-\gamma}\,\frac{T_c(1+\lambda)}{eD}\left(1 + 3.3\frac{T_c}{\omega_{log}} - 4.8\left(\frac{T_c}{\omega_{log}}\right)^2 \ln\left(\frac{\omega_{log}}{T_c}\right)\right), \quad (4)$$

where the diffusion coefficient $D = 1/3\ V_F^2 \times \tau$, and $\tau$ – the electron scattering time on impurities and defects which is unknown.

The difficulty in applying these formulas is that they require knowledge of the averaged Fermi velocity, electron-phonon coupling parameter λ, and the electron mean free path. Moreover, the dependence on these parameters is very strong, leading to large uncertainties in the calculated value of $B_{c2}(0)$. The best fits of the experimental data for superhydrides in the clean and dirty limits and were performed earlier [16] and lead to

$$B_{c2}^{cl}(0) = 1.855 \cdot 10^{-4} T_c^2(1+\lambda)^2\left(1 + 1.44\frac{T_C}{\omega_{log}}\right), \quad (3a)$$

$$B_{c2}^{di}(0) = 0.126\, T_c(1+\lambda)\left(1 + 3.3\frac{T_c}{\omega_{log}} - 4.8\left(\frac{T_c}{\omega_{log}}\right)^2 \ln\left(\frac{\omega_{log}}{T_c}\right)\right). \quad (4a)$$

Using these eqs. (3a), (4a), the data in Table 2, and the experimental $T_c$ values, we obtain that $B_{c2}^{cl}(0)$ = 77 T and 142 T for $YH_6$ and $YH_9$ at 200 GPa, respectively, and $B_{c2}^{di}(0)$ = 96 T and 133 T for the same hydrides at the same pressure. These values are in a good agreement with the density functional theory (DFT) calculations performed within the Migdal-Eliashberg (ME) formalism [35]. Comparing with experimental values in Table 1 (max $B_{c2}(0)$ = 115–157 T for $YH_6$ and 100–120 T for $YH_9$), the upper critical fields of the synthesized yttrium hydrides are clearly better described by the dirty- limit formulas.

*II. Radio-frequency transmission measurements of $YH_6$*

The Meissner effect, AC susceptibility, or NV-center magnetometry has never previously been studied in yttrium superhydrides. Such studies exist for sulfur hydrides ($H_3S$) [36,37], cerium hydrides ($CeH_9$) [38] and lanthanum polyhydrides ($LaH_{9-12}$) [18,36]. Here we fill this gap for *bcc*-$YH_6$ using our recently developed high-frequency AC susceptibility technique in a radio-frequency transmission configuration. Unlike previous studies, we used Pt/Ta microcoils (thickness is ~1 µm) sputtered onto the diamond anvils, forming Lenz lenses[39] with a minimum diameter of about 30 µm, a multi-turn (N = 20) inductor and RF receiver, and a higher operating frequency of 500 kHz to 5 MHz (Figure 2).

We have used the contactless radio-frequency transmission method in numerous studies of the La-H [18], La-Ce-H [17], and La-Sc-H [40] superhydride systems , as well as to the study of Nd-doped 327-type nickelates [41], Bi-based cuprates (e.g., Bi-2223 [42]) and low-temperature superconductors [17,19]. Here we employed an improved version of the method with a multi-turn emitter and receiver coils, which resulted in a signal amplification by 1-2 orders of magnitude compared with earlier studies. The radio-frequency experiment was subsequently repeated without XRD analysis, in a DAC Y4 at 140-148 GPa with the same result, but a lower $T_c$ of 211 K (Supporting Figure S15).

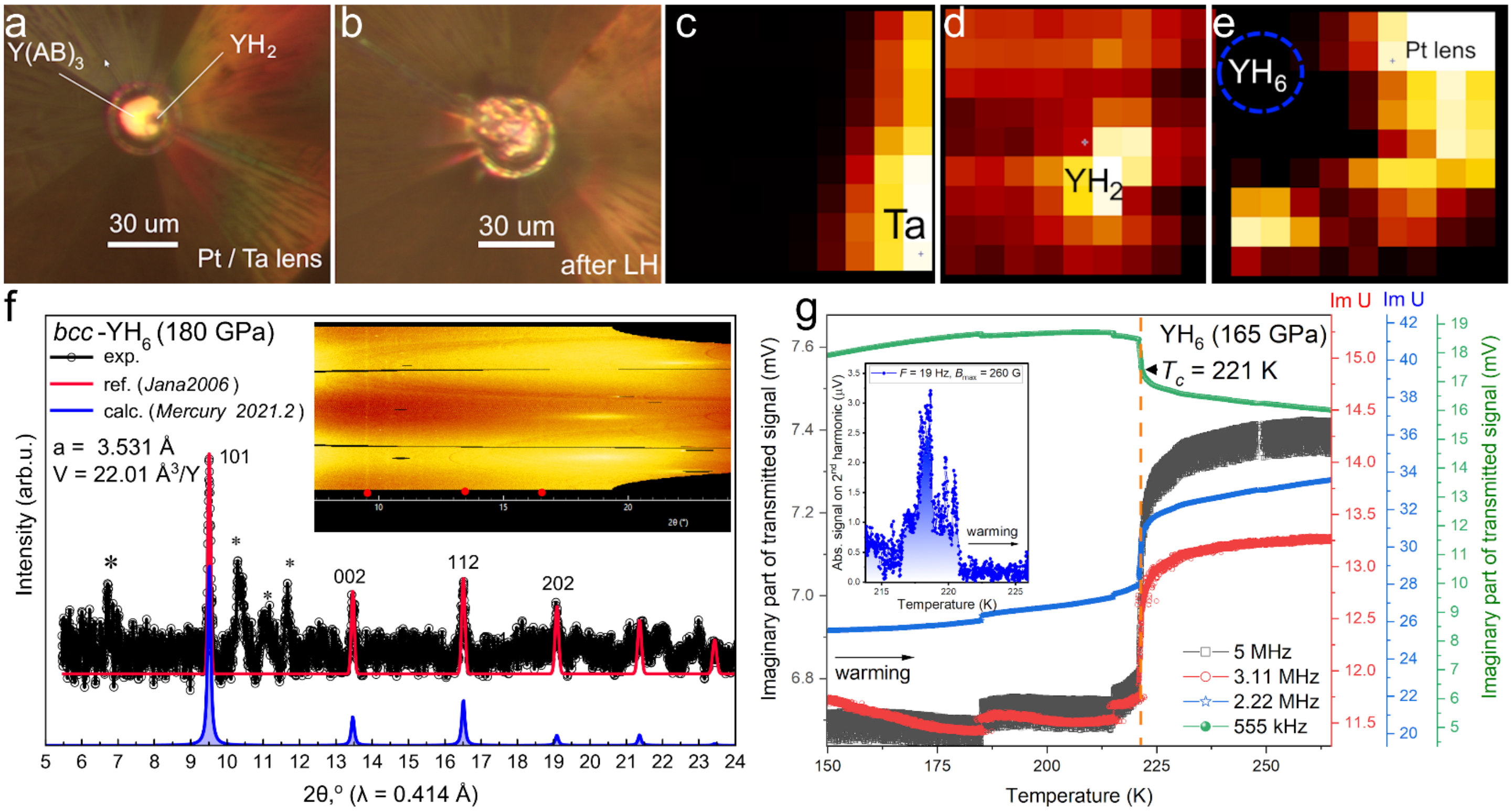

**Figure 2.** X-ray diffraction and radio-frequency transmission study of $YH_6$ synthesized at 180 GPa in a DAC Y3 from a $YH_2$/$Y(AB)_3$ mixture. (a, b) Optical micrographs of the sample chamber before and after laser heating (LH), respectively, showing the Pt/Ta Lenz lens. (c–e) 2D X-ray diffraction mapping demonstrating the distribution of Ta (c), unreacted $YH_2$ (d), and the synthesized $YH_6$ phase (e, highlighted by a dashed blue circle) near the culet center. (f) Integrated X-ray diffraction pattern of the cubic *bcc*-$YH_6$ phase at 180 GPa (λ = 0.414 Å). Experimental data are represented by black circles, the refinement profile (Jana2006) by a red line, and the calculated spectrum (Mercury 2021.2) by a blue line. The determined lattice parameter is $a$ = 3.531 Å ($V$ = 22.01 Å³/Y). Asterisks denote unexplained diffraction peaks. The inset displays the raw 2D X-ray diffraction image ("cake"). (g) Temperature dependence of the imaginary part of the transmitted radio-frequency signal measured at various frequencies (from 555 kHz to 5 MHz). The measurements were performed at 165 GPa due to a pressure drop following the laser heating. A clear superconducting transition is observed at a critical temperature $T_c$ = 221 K. The inset illustrates the absolute signal on the 2nd harmonic of the modulating field (19 Hz, 260 Gauss).

The synthesis of $YH_6$ followed an original route first proposed and tested by us for $RbH_9$ and $CsH_7$ [43]. Anhydrous $YCl_3$ was ball-milled with $LiNH_2BH_3$ under argon for 10 minutes, yielding $Y(AB)_3$ mixed with LiCl. The resulting $Y(AB)_3$/LiCl mixture was loaded into the chamber of DAC Y3 through a hole in the $TaW_{10\%}/Ta_2O_5$ composite insulating gasket; the high-pressure chamber diameter was 40–45 μm (Figure 2a). Under laser-assisted high-temperature conditions, the yttrium amidoborane decomposes to the target hydride:

$$YCl_3 + 3LiNH_2BH_3 \rightarrow Y(NH_2BH_3)_3 \xrightarrow{LH} YH_6 + 3BNH_x + LiCl + \ldots \qquad (5)$$

After compression to 180 GPa, the mixture was heated with a pulsed IR laser (1064 nm) at the BL10XU station of the SPring-8 synchrotron source. The heating results were monitored via the diffraction pattern and by mapping the high-pressure chamber area (Figure 2c, d, e). Because $Y(AB)_3$ is transparent, we used a $YH_{2-x}$ yttrium particle and the edge of a Pt/Ta Lenz lens (Figure 2a, b) as laser heating targets. After heating, the pressure in the DAC Y3 dropped to 165 GPa.

A weak diffraction signal of the synthesized *bcc*-$YH_6$ ($a$ = 3.531 Å, $V$ = 22.01 Å$^3$/Y, Figure 2f) was detected after laser heating in the culet center. Its low intensity is due to the small culet size, the thin layer of the $Y(AB)_3$ precursor at 180 GPa, and the weaker X-ray scattering form factor of Y compared to La. Despite its small amount, the synthesized $YH_6$ produces a very pronounced step-like signal (of a few mV) at the superconducting transition in the RF transmission at all studied frequencies near 221-225 K during warming and cooling cycles (see also Supporting Figure S14), as well as in the 2$^{nd}$ harmonic of the external AC magnetic field (19 Hz, $B_{max}$ = 260 G, see Figure 2g). Thus, we detected a signature of the susceptibility change in the synthesized $YH_6$, which is one of the key hallmarks of superconductivity.

### *III. Effect of palladium and aluminum on superconductivity in yttrium hydrides*

Palladium dissolves large amounts of hydrogen [44] forming the superconducting hydride $PdH_{1-x}$ [45]; combined with its low hydrogen-binding energy, this makes Pd an effective hydrogenation catalyst [46] and a hydrogen-transparent protective coating for reactive metals [47]. And finally, Pd is known with its strongly correlated multielectron system that is close to the Stoner instability [48].

Therefore, the question of how palladium influences the superconductivity of superhydrides and whether it can be used in their synthesis is quite topical. This issue became particularly interesting to elucidate following the publication (and subsequent retraction) of the paper by R. Dias, A. Salamat et al. [49] on the synthesis of an unknown yttrium hydride with $T_c$ = 262 K using a thin palladium layer deposited on yttrium as catalyst. Under intense laser heating to 1000–2000 K, such a layer can diffuse into the yttrium sample and form stable intermetallics (YPd, $Y_3Pd$, etc.) [50]; moreover, a thin Pd film, reacting with hydrogen at high pressure is likely to change its mechanical properties and crack under high-pressure high-temperature conditions [51].

To investigate the effect of 25 at% of palladium in an yttrium hydride sample, we synthesized *Pnma*-$Y_3Pd$ by arc melting at ~3000°C from a suitable mixture of yttrium and palladium metal pieces (~ 3 g). Powder XRD data (Figure 3b) confirms a sufficiently pure *Pnma*-$Y_3Pd$ product. This sample was then used to synthesize polyhydrides in a DAC Y5 (50 μm culet) at 174 GPa (Figure 3a) with a standard $NH_3BH_3$ hydrogen source and pulsed IR laser heating (1.06 μm). The electrical resistance measurements of the resulting $(Y_3Pd)H_x$ after two laser heating stages (LH1, LH2, Figure 3c) indicate that the $(Y_3Pd)H_x$ does not exhibit superconducting properties above 120 K (see also Supporting Figure S10). Application of the Bloch-Grüneisen fit of $R(T)$ yields a relatively high Debye temperature ($\theta_D$ = 980–1725 K), characteristic of polyhydrides. These results indicate a pronounced suppression of superconductivity in yttrium hydrides by palladium.

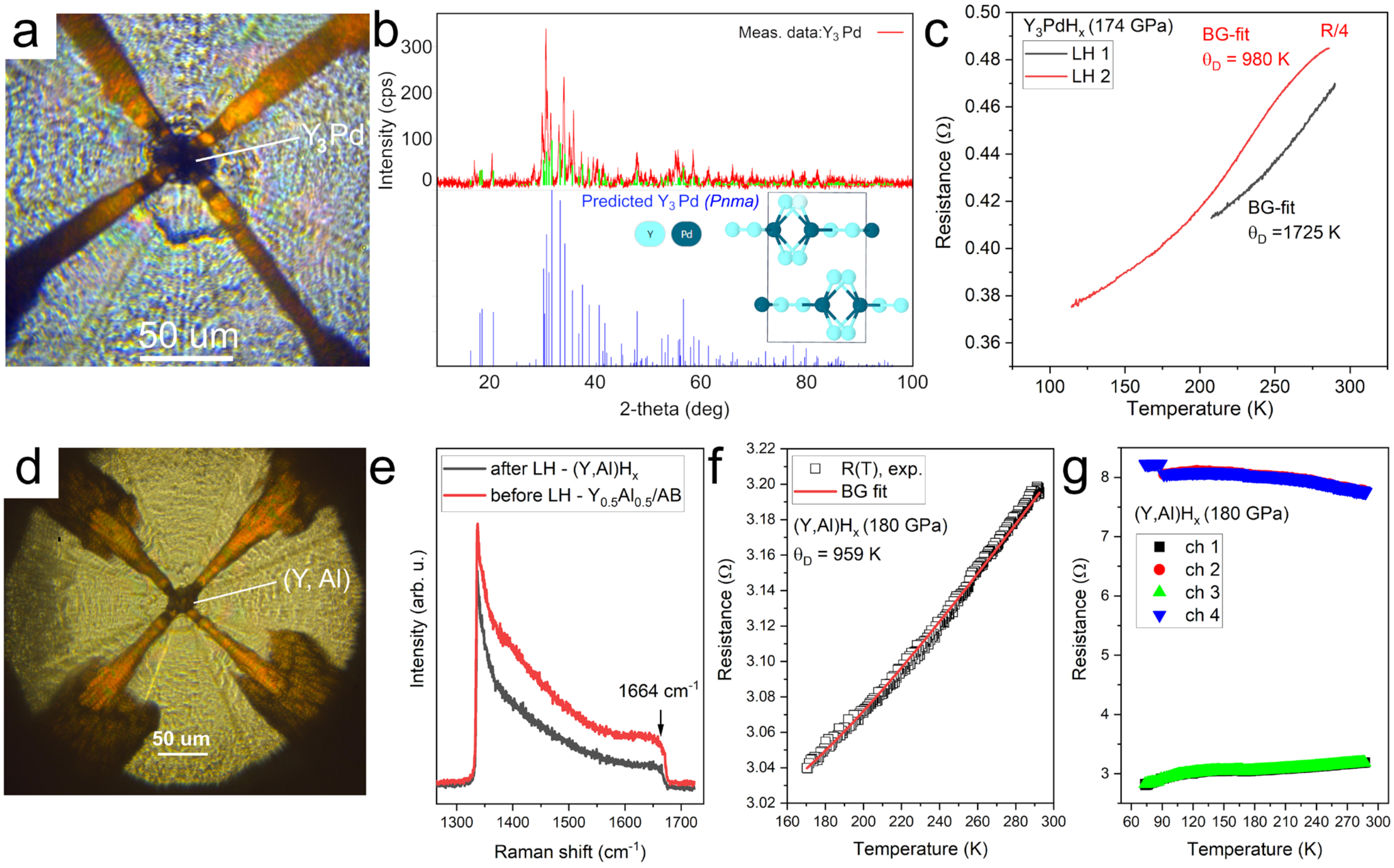

**Figure 3.** Absence of high-temperature superconductivity in the Y-Pd(25 at%)-H and Y-Al(50 at%)-H systems at 174-180 GPa. (a) Photo of the $Y_3PdH_x$ sample obtained after laser heating of $Y_3Pd/NH_3BH_3$ in DAC Y5. (b) Powder X-ray diffraction pattern (Cu $K_a$) of the precursor for synthesis, *Pnma*-$Y_3Pd$, and the expected diffraction pattern, as well as the structure of this intermetallic compound. (c) Temperature dependence of electrical resistance of the obtained $Y_3PdH_x$ after the first and second laser heatings (LH1, LH2) in the range from 290 to 120 K, as well as fitting by the Bloch-Grüneisen formula (eq. 1). The samples demonstrate metallic behavior with the absence of high-$T_c$ superconductivity. (d) Photo of the $(Y_{0.5}, Al_{0.5})H_x$ sample in DAC Y7 after laser heating at 180 GPa. (e) Raman spectroscopy of diamond in the cullet region before and after laser heating, showing little pressure change in DAC Y7. (f) Temperature dependence of electrical resistance of $(Y, Al)H_x$ (channel 1) in the range of 170-290 K and the Bloch-Grüneisen fit, indicating a high Debye temperature and probable polyhydride synthesis. (g) Temperature dependence of electrical resistance in all four channels (ch1-4) of the van der Pauw scheme in the range from 78 K to 290 K for DAC Y7. No high-temperature superconductivity is observed.

An additional experiment used a 50–100 nm thick palladium film deposited on an yttrium sample shaped as a ring (Y/Pd) on a 30 μm culet (Supporting Figure S9), the center of which was filled with $NH_3BH_3$ to increase the hydrogen content of the target hydride. The synthesis was carried out in a DAC Y6 equipped with eight electrodes and a Lenz lens, at 180 GPa (Supporting Figure S9a). Upon heating with a pulsed IR laser (1.06 μm), the interior of the Y/Pd ring was filled with expanding hydride material. Despite repeated laser heating, superconductivity was detected only at low temperatures: below 20 K by radio-frequency methods using the Lenz lens as a detector (Supporting Figure S11d), and below 13-15 K by direct resistance measurements (Supporting Figures S11a-c). Thus, in this case too, the presence of palladium significantly worsened the superconducting properties of yttrium hydrides.

A DFT calculation for a model structure $YPdH_{12}$, obtained from the *Im*$\bar{3}$*m*-$Y_2H_{12}$ by regular substitution of yttrium with palladium at 180 GPa, on a coarse q-mesh (2×2×2) yields a nominally high $T_c$(AD) ≈ 210 K (AD – stands for Allen-Dynes formula [22], Supporting Figure S19), but an unrealistically large λ = 3.72, exceeding the fundamental limit of 3.69 for electron-phonon interaction with Einstein phonons [52]. This indicates that the compound is dynamically unstable in the harmonic approximation, and cannot provide such a high $T_c$. However, in the experiments, regular substitution of half of the yttrium positions by palladium does not occur. The substitution is most likely random, as in most ternary hydrides [53]. Summarizing the experiment with DACs Y5 and Y6, we observe no positive effect of palladium on superconductivity in yttrium hydrides, in contrast to the claims of Ref. [49].

Aluminum hydrides were studied at the very beginning of the "hydride era" (~2008); the only aluminum hydride synthesized to date, $AlH_3$, is expected to superconduct only below 2 K [54]. Given these poor prospects, the outcome of the experiment in the DAC Y7 sample is not surprising (Figure 3d). DAC Y7 was loaded with a 1:1 Y-Al alloy, possibly also containing the intermetallic AlY, prepared by sintering of yttrium and aluminum in a muffle furnace in a sealed crucible for several days at 1400 ºC. After laser heating of the sample with $NH_3BH_3$ at 180 GPa (Figure 3e), no traces of high-$T_c$ superconductivity were found in the range from 78 K and 290 K. Thus, aluminum, like palladium, is a pronounced suppressor of superconductivity in hydrides.

Aluminum has previously been used to dope the La-H system at concentrations up to 30 at% at 150-160 GPa [55]. The superconducting onset temperatures achieved were 180-223 K, at least 25 K below the maximum for $LaH_{10}$ (250 K [12]). The superconducting transitions broadened, did not reach zero resistance above 200 K, and when the aluminum content reached 30 at%, superconductivity in the samples completely disappeared. Our result for 50 at% aluminum doping, with its sharply negative effect on superconductivity, is fully consistent with Ref. [55].

*IV. Theoretical analysis of anharmonic effects*

The Y–H superhydride phases have been investigated theoretically through first-principles calculations by several groups in recent years [1,2,35,56-62]. It is now well established that high-$T_c$ superconductivity in these compounds arises from strong and relatively uniform electron-phonon (el-ph) coupling involving both the hydrogen and yttrium sublattices. The hydrogen-rich sodalite-like clathrate geometry gives rise to high-frequency hydrogen vibrational modes and a substantial contribution of hydrogen to electronic states in the band structure near the Fermi level. At the same time, yttrium states also contribute to the Fermi surface, while strongly hybridized Y–H interactions make the superconducting gap nearly isotropic across the Fermi surface.

Earlier studies, conducted before the widespread adoption of anharmonic phonon methods[63,64], found anharmonic effects to be relatively modest in these compounds [2]. This contrasts with the pronounced quantum anharmonic effects established in other superhydrides, such as $H_3S$ [65] and $LaH_{10}$ [66], and more recently in $SrPdH_3$ [67]. More recent investigations have suggested that anharmonicity may help explain the discrepancies between theoretical predictions and experimental measurements in $YH_6$ and $YH_9$ [61], for which $T_c$ is typically overestimated by at least 20 K. Nevertheless, a comprehensive and systematic analysis of quantum anharmonic effects in $YH_6$, $YH_9$, and $YH_{10}$, performed within a unified methodological framework and based on advanced ab initio techniques such as the SSCHA, is still lacking in the literature.

In this vein, we performed density-functional perturbation theory (DFPT) [68] calculations for all these phases at experimentally accessible pressures and incorporated quantum anharmonic lattice effects using SSCHA. The SSCHA-corrected dynamical matrices were then used to recompute the el-ph coupling and estimate $T_c$ by solving the isotropic Migdal–Eliashberg equations within the full-bandwidth approximation (FBW) [59], as implemented in the IsoME code [69]. We also rescaled the Eliashberg spectral function by the density of states (DOS) at the Fermi level ($N_F$), computed using the accurate tetrahedron method, to correct for the inaccurate value of the Fermi-level DOS obtained on coarser grids with the smearing method, following Ref. [70,71].

Figure 4 and Table 2 summarizes our theoretical findings. First, all investigated Y–H phases are dynamically stable at the pressures considered. In particular, quantum anharmonic lattice effects stabilize $YH_9$ at 200 GPa and $YH_{10}$ at 250 GPa by hardening the soft harmonic instabilities and shifting them to real phonon frequencies. From the point of view of the stability analysis of metallic state ($\xi$-criterion, Supporting Equation S3) [52] with respect to the strong electron-phonon interaction, all yttrium hydrides listed in Table 2 are stable in the anharmonic approximation ($\xi < 0.5$) even at the smallest applied broadening parameter $\sigma = 0.02$ Ry. However, if we look at the harmonic approximation for $YH_{10}$ at 250 GPa [1,2,53], the stability parameter already exceeds 0.52 in the σ-interval, where $T_c \geq 290$-300 K is obtained – an independent indication that the room-temperature superconductivity is unlikely to be realized in $YH_{10}$.

**Table 2.** Computed superconducting properties of $YH_6$, $YH_9$, and $YH_{10}$. Here $P$ is the pressure considered, and $N_F$ is the total DOS at the Fermi level in states/eV/formula unit. The quantities λ, $\omega_{log}$, and $T_c$ are, respectively, the total el-ph coupling parameter, the logarithmic average phonon frequency, and the superconducting critical temperature obtained within isotropic FBW-ME theory; ξ – is the stability criterion. The labels "HA", "SSCHA", and "exp" indicate quantities computed within the harmonic and anharmonic approximations and the experimental values reported in the present work. $T_c$ was obtained using $\mu^*$(AD) = 0.11.

| Compound | P [GPa] | $N_F$ [eV$^{-1}$] | λ | | $\omega_{log}$ [meV] | | $T_c$ [K] | | | |
|---|---|---|---|---|---|---|---|---|---|---|
| | | | HA | SSCHA | HA | SSCHA | HA | SSCHA (ξ) | SCDFT | exp |
| $Im\bar{3}m$- $YH_6$ | 200 | 0.67 | 1.93 | 1.65 | 110 | 108 | 224 | 198 (0.33) | 175 | 224-226 |
| $P6_3/mmc$- $YH_9$ | 200 | 1.10 | 6.50 | 2.20 | 38 | 82 | 277 | 238 (0.41) | 191 | 230-243 |
| $Fm\bar{3}m$- $YH_{10}$ | 250 | 0.75 | 3.33 | 2.17 | 77 | 109 | 283 | 260 (0.49) | 221 | < 260-270* |

* The estimation is based on $T_c$ of substituted ternary hydrides $(Y,X)H_{10}$.

Before including anharmonic corrections, both compounds display a continuous phonon spectrum. A small fraction of hydrogen-derived phonon modes extends from energies as low as 10 meV up to 250 meV, while the yttrium modes are confined to a narrower energy window between 0 and 40 meV. Notably, both compounds also exhibit a broad manifold of soft modes between approximately 50 and 100 meV, which partially accounts for the unrealistically large el-ph coupling strength, exceeding the commonly accepted regime $\lambda < 3$ for dynamically stable structures.

When anharmonicity is included within the SSCHA, these soft modes are stabilized and the low-frequency hydrogen modes vanish. This leads to a clear separation between hydrogen and yttrium vibrations, with a phonon gap opening between approximately 50 and 100 meV. As a result of this strong renormalization of the phonon spectrum, λ is reduced to approximately 2.2 for $YH_9$ and 1.8 for $YH_{10}$ and $\omega_{log}$ is increased to 82 meV for $YH_9$ and 109 meV for $YH_{10}$.

For $YH_6$ at 200 GPa, the changes are less pronounced, since this structure is dynamically stable at the harmonic level and exhibits a clear separation between the hydrogen and yttrium phonon manifolds. Nevertheless, the anharmonic treatment hardens the soft hydrogen phonon modes above 100 meV and shifts the phonons above 200 meV to higher frequencies. The suppression of these soft modes also reduces λ, from 1.9 to 1.7.

The superconducting properties computed within the harmonic and anharmonic approximations are summarized in Table 1, where the resulting $T_c$ values are compared with the available experimental data. Notably, we obtain excellent agreement with experiment, demonstrating the key importance of quantum anharmonic lattice effects in this class of superhydrides as well. For $YH_9$, for example, the discrepancy with the experimental $T_c$ is reduced from 16% to about 3% once anharmonic effects are included.

In addition, we find that, even within the harmonic approximation, substantially improved agreement with the experimental $T_c$ can be achieved when the following key ingredients are included in the evaluation of $T_c$. The first is the use of the appropriate electronic energy scale for the Coulomb pseudopotential in the solution of the ME equations, as discussed in Ref. [72]. This scale differs from the one entering the Allen-Dynes formulation, which is commonly adopted in the literature.

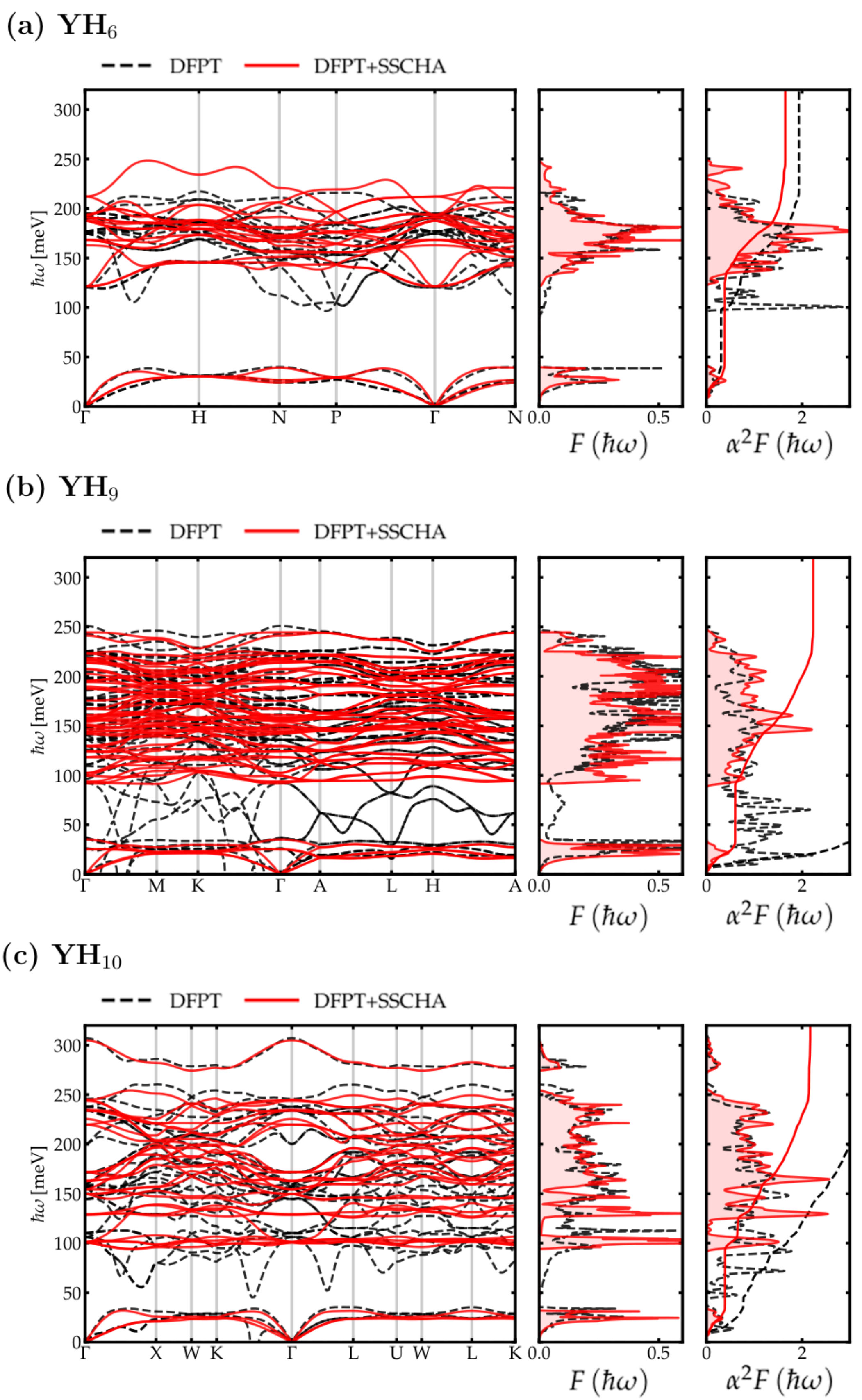

**Figure 4.** Phonon spectra, el-ph interactions, and superconducting properties of $YH_6$, $YH_9$, and $YH_{10}$ within the harmonic approximation (black dashed lines) and including anharmonic quantum lattice effects (red solid lines). For each compound, we show the phonon dispersion along high-symmetry paths in the Brillouin zone, the total phonon DOS $F(\hbar\omega)$, and the Eliashberg spectral function $\alpha^2 F(\hbar\omega)$, together with the cumulative el-ph coupling parameter $\lambda$.

The second ingredient is going beyond the constant-DOS approximation in Eliashberg theory, thereby accounting for electronic states away from the Fermi level in the superconducting pairing and updating the chemical potential self-consistently. As demonstrated in Ref. [59], this effect can modify $T_c$ by several kelvin and mitigates inaccuracies associated with the integration of the double-delta functions on coarse grids [59,70]. Taken together, these ingredients yield $T_c$ = 224 K for $YH_6$ at 200 GPa within the harmonic approximation, less than 10 K away from the experimental value reported at 195 GPa in the present work.

The importance of appropriately rescaling the Coulomb pseudopotential together with the use of the FBW formulation of the ME equations is also evident for $YH_{10}$. Even within the harmonic approximation, we obtain $T_c$ = 283 K, substantially lower than the above-room-temperature value of 310 K previously predicted [1,2]. The inclusion of anharmonic effects further reduces $T_c$ to 260 K. Thus, our results revise the expected $T_c$ of $YH_{10}$, should this phase eventually be synthesized experimentally, and reinforce the need to continue searching for new superhydrides capable of exceeding room temperature superconductivity at ultra-high pressures.

The Coulomb effect was evaluated from first principles by the use of the Superconducting Density Functional Theory **(**SCDFT) which enables us to calculate superconducting critical temperature ($T_c$) without the empirical $\mu^*$ parameter. Results presented in the Table 2 reveal a significant improvement in predicting the superconducting critical temperatures ($T_c$) of yttrium hydrides when using the SPG2020 [73] functional compared to the older LM2005 model [74] (see Supporting Tables S2, S3) For both $YH_9$ and $YH_6$, SPG2020 yields higher $T_c$ values: 191 K and 175 K, respectively, which align more closely with the experimental results than the underestimated values from LM2005 model. Physically, this improvement is attributed to the formulation of SPG2020, which treats the renormalization of the electronic Green's function more accurately, thereby correcting, though found not perfectly [75], the systematic underestimation observed in earlier models. Nevertheless, the discrepancy of 40-50 K with the experimental $T_c$ (Table 2) clearly indicates the need to improve the existing functionalities for SCDFT calculations.

*V. Can $YH_{10}$ be a room-temperature superconductor?*

According to our results, the answer is no. Its $T_c$ is likely 260–270 K (221 K) according to the FBW-Eliashberg (SCDFT). In fact, this was already noted in several early papers (for instance, Ref. [62]), when there was a buzz surrounding hydride superconductivity. Experiments with solid solutions of $(La,Y)H_{10}$ directly showed that $T_c$ increases only slightly, within just 3 K, when about 50% of La is replaced by Y [13].

More recently, another experiment on the synthesis of $(Y, Ce)H_{10}$ containing 33 at% of cerium, reported a maximum $T_c$ of 201 K at around 150 GPa [76]. The critical temperature of $(Y, Ce)H_{10}$ likely decreases to 180–190 K when $T_c(P)$ is extrapolated to 250 GPa with $dT_c/dP$ = –0.3 K/GPa [76]. As we know, 1 at% of Ce suppresses superconductivity in $(La,Ce)H_{10}$ by 1.25–2.5 K [53,77-79], hence 33 at % of Ce will suppress the $T_c$ by 41-82 K. Assuming similar linear suppression of $T_c$ by Ce in yttrium superhydrides, a simple calculation gives $T_c(YH_{10})$ = 242–283 K at 150 GPa, with a mean value of 262 K. This is an upper estimate, because stabilizing *fcc* $YH_{10}$ at 150 GPa has not been achieved to date. At its stabilization pressure of 250 GPa, $T_c$ in $(Y,Ce)H_{10}$ decreases to 180-190 K, with a coefficient $dT_c/dP \approx -0.3$ K/GPa. This gives a maximum $T_c(YH_{10})$ of 262 K at 250 GPa, in good agreement with our SSCHA calculations (Table 2). We emphasize that both assumptions are strong – impurity pair-breaking need not remain linear over such a wide concentration range, and the electronic state of Ce may differ between the La and Y host lattices – so this estimate should be regarded as an independent consistency check on, rather than a substitute for, our SSCHA result.

To gain insight into the stabilization pressure of $YH_{10}$, we used the MTPs trained at 250 GPa to evaluate the SSCHA-corrected phonon dispersion at 200 GPa. The resulting dispersion exhibits no imaginary modes (Supporting Figure S20), suggesting that quantum anharmonic lattice effects may stabilize $YH_{10}$ even 200 GPa. However, a quantitative determination of the stability boundary would require training pressure-specific MTPs at progressively lower pressures until the onset of imaginary phonon modes is identified.

Thus, despite numerous discussions [1,2] regarding room-temperature superconductivity in $YH_{10}$, even if this compound were synthesized, its critical temperature is unlikely to exceed 260–270 K.

## Conclusions

Using contact electrotransport and contactless radio-frequency transmission techniques, we studied superconductivity in the $YH_6$ ($T_c \approx 218$ K at 195 GPa) and $YH_9$ ($T_c \approx 235$–236 K at 205–213 GPa) superhydrides, and reported the first direct observation of the RF response of $YH_6$. Pulsed-field measurements up to 60 T reveal that $B_{c2}(T)$ is consistent with the dirty-limit behavior of BCS-ME superconductors. Doping of yttrium hydrides with palladium and aluminum strongly suppresses of high-$T_c$ superconductivity in both cases, with no transitions detected above 120 K (Y-Pd) or 78 K (Y-Al). On the theoretical side, we show that quantum anharmonic lattice effects incorporated via SSCHA are indispensable for quantitative $T_c$ prediction. Applying this framework to cubic $YH_{10}$ at 250 GPa revises its predicted $T_c$ downward from the widely cited ~310–326 K to approximately 260–270 K, ruling out room-temperature superconductivity in known binary yttrium superhydrides.

## Data Availability

Authors declare that the main data supporting our findings of this study are contained within the paper and Supporting Information. All relevant data are available from the corresponding authors upon request.

## Authors Contributions

D.V.S., I.A.T., and D.Z. performed the high-pressure experiments. P.N.F., F.J., R.A. and C.H. carried out theoretical calculations. A.V.S., T.H, and V.M.P. executed transport measurements in magnetic fields. K.S.P. synthesized the $Y_3Pd$ intermetallics and AlY alloy. B.I.M. prepared lenses for experiments with radio-frequency transmission experiment. F.J. performed the SSCHA calculations. P.N.F. performed the calculations of phonons, electron-phonon coupling, and $T_c$ within the ME formalism. R.A. performed the calculations of $T_c$ within the SCDFT formalism. All authors contributed to the analysis and interpretation of the results. D.V.S., P.N.F., V.M.P., T.H. and V.V.S. contributed to writing and revising the manuscript. D.V.S. and V.V.S. supervised the project. All authors discussed the results presented in the manuscript and provided comments.

## Acknowledgments

D.V.S. and D.Z. are grateful for the financial support from HPSTAR. This work was supported by the National Key Research and Development Program of China (grant 2023YFA1608900, subproject 2023YFA1608903). V.V.S. acknowledges financial support from the Shanghai Science and Technology Committee, China (No. 22JC1410300) and Shanghai Key Laboratory of Materials Frontier Research in Extreme Environments (MFree), China (No. 22dz2260800). Computational resources were provided by the Austria Scientific Computing clusters (ASC4 and ASC5). P.N.F. acknowledges support from the Austrian Science Fund (FWF) under project DOI 10.55776/ESP8588124. A. V. S. and I. A. T thank Russian Science Foundation, grant RSF 26-12-00447, for the financial support of the research. D.V.S. is grateful for financial support from the EMFL-ISABEL project. This work was supported by HLD-HZDR, member of the European Magnetic Field Laboratory (EMFL).

SUPPORTING INFORMATION

# Yttrium Superhydrides Revisited:

# Advanced Experimental and Theoretical Studies of $YH_6$, $YH_9$ and $YH_{10}$

Dmitrii V. Semenok,[1,†,*] Pedro N. Ferreira,[2,†,*] Di Zhou,[1,†] Fabian Jőbstl,[3] Andrey V. Sadakov,[4] Kirill S. Pervakov,[4] Burkhan I. Massalimov,[4] Toni Helm,[5] Ryosuke Akashi [6], Vladimir M. Pudalov,[4,9] Viktor V. Struzhkin,[7,8] Christoph Heil,[2] and Ivan A. Troyan[7,*]

[1] Center for High Pressure Science & Technology Advanced Research, Bldg. 8E, ZPark, 10 Xibeiwang East Rd, Haidian District, Beijing, 100193, China

[2] Institute of Theoretical and Computational Physics, Graz University of Technology, NAWI Graz, 8010, Graz, Austria

[3] University of Vienna, Faculty of Physics, Kolingasse 14-16, Vienna, Austria

[4] V. L. Ginzburg Center for High-Temperature Superconductivity and Quantum Materials, 53 Leninsky Prospekt, building 10, Moscow 119991, Russia

[5] Hochfeld-Magnetlabor Dresden (HLD-EMFL) and Würzburg-Dresden Cluster of Excellence, Helmholtz-Zentrum Dresden-Rossendorf (HZDR), Bautzner Landstrase 400, Dresden 01328, Germany

[6] Department of Applied Physics and Physico-Informatics, Keio University, 3-14-1 Hiyoshi, Yokohama, Kanagawa 223-0061, Japan

[7] Shanghai Key Laboratory of Material Frontiers Research in Extreme Environments (MFree), Shanghai Advanced Research in Physical Sciences (SHARPS), 68 Huatuo Rd, Bldg 3 Pudong, Shanghai 201203, China

[8] Center for High Pressure Science & Technology Advanced Research, 1690 Cailun Rd, Bldg 6, Pudong, Shanghai 201203, China

[9] National Research University Higher School of Economics, Moscow 101000, Russia

*Corresponding authors: Dmitrii V. Semenok (dmitrii.semenok@hpstar.ac.cn), Pedro N. Ferreira (nunesferreira@tugraz.at), and Ivan A. Troyan (itrojan@sharps.ac.cn)

† These authors contributed equally

## Content

# I. Methods

*Experiment*

Y-Pd and Y-Al alloys were prepared by arc melting of metals in the desired atomic ratio, in an argon atmosphere at a temperature of about 3000-4000 ºC (MAM-1, Edmund Bühler GmbH). Ammonia borane (AB) for the synthesis was purified by vacuum sublimation at 50 ºC from commercial $NH_3BH_3$ (AB, Sigma Aldrich, Figure S1). Laser heating of metal samples in the AB media at high pressure was carried out through a series of ~0.3 s infrared laser pulses (λ = 1.06 µm), in which the sample temperature increased to at least 1000-1500 K leading to the decomposition of AB. The pressure in DACs was measured at the edge of the diamond Raman signal using the Eremets scale (DACs Y1-2, Y5-7) [1] and the Akahama scale (DACs Y3-4, Y6) [2]. It is important to emphasize that the non-uniform pressure distribution across the culet results in an error in pressure determination of the order of ± 5 GPa. Moreover, the pressure in DACs typically increases as the sample cools to cryogenic temperatures. For electrical resistance measurements, a four Ta/Au electrode van der Pauw scheme was used, the current amplitude was from 0.1 to 1 mA, the frequency, depending on the purpose, varied from a DC up to 7.77 kHz for transport measurements in pulsed magnetic fields.

Crystal structure of $YH_x$ polyhydrides synthesized in DACs Y3, and Y6 was studied using the synchrotron X-ray diffraction (XRD) at the BL10XU beamline with a beam of wavelength of 0.414 Å at the SPring-8 Synchrotron Research Facility at 300 K. Crystal structure of $Y_3Pd$ intermetallic was studied using Rigaku MiniFlex 600 powder X-ray diffractometer. Mapping of the DACs Y3, Y6 samples' center was carried out on a 3×3, 4×4 and 5×5 grids with a step of 5-8 µm, accumulating time was 2 seconds per image with 10-30 images averaging. The experimental XRD images were integrated and analyzed using the Dioptas 0.7 software package [3]. To fit the diffraction patterns and obtain the cell parameter, we analyzed the data using Mercury 2021.2.0 [4] and Jana2006 software [5], employing the Le Bail method [6]. Analysis of the spatial distribution of hydride phases in the sample was performed using the Dioptas 0.7 program [7].

Resistance measurements in pulsed magnetic field were performed in a four-contact van der Pauw scheme using an alternating current of 2 mA (RMS) with a frequency of 7.77 kHz. To analyze the magnetic phase diagram, we used the real part of the impedance ($Z = U/I$) corrected by the phase factor so that in the superconducting state Re Z = 0. The duration of the magnetic field pulses was 150 ms, the maximum field amplitude was 68 Tesla. To control the temperature, a Cernox thermometer was glued to the gasket with thermoconductive glue. To stabilize the sample temperature, an external heater made of Ni-Cr wire of 0.8 m long and 50 µm in diameter, was used.

The radio-frequency (RF) measurements, described in our previous publications [8-11], were performed in DACs Y3, Y4, and Y6, which had two symmetrical three-stage Lenz lenses consisting of a bimetallic Ta/Pt or Ta/Au layers deposited on diamond anvils using a JEOL IB-09010CP Cross Section Polisher. The argon ion beam energy was 7.5-8 keV, and the deposition time was 5-7 minutes for each metal layer. The central part of the culet was then etched using a focused Ga beam (Ga FIB, Thermo Fisher Scientific) until it reached the diamond surface. Then, a 1 µm thick Y or Y/Pd layer was deposited on the anvil using the same Cross Section Polisher, and Y or Pd target. The deposition time was 30 minutes at an ion beam energy of 7.5 keV. Finally, re-cutting using Ga FIB allowed the formation of a Lenz lens topology and an annular target made of Y or Y/Pd. In the case of the DAC Y4 sample, as a starting precursor, we used an yttrium microparticle loaded with a tungsten needle. A detailed description of all the experiments carried out is given in Table S1.

We performed measurements of the transmission of high-frequency signals through the Lenz lens system in the DACs Y3, Y4, and Y6 in a circuit with two lock-in amplifiers. We used a MFLI (4 generators and 4 demodulators at the same time) or SR844 lock-in amplifier as high-frequency signal generator, as well as receiver to measure the transmitted signal at the same frequency. Time constant was 0.3-1 ms. We placed the sample in a low-frequency magnetic field (19 Hz, maximum induction is 60-260 Gauss), generated by a special solenoid. The current in the solenoid was created by an SR830 lock-in amplifier, working in tandem with a Yamaha PX10 power amplifier. The envelope of the high-frequency signal near the superconducting transition temperature contains even harmonics, of which the strongest, the $2^{nd}$ harmonic, was detected, thus serving as an additional marker of the superconducting transition sensitive to the magnetic field.

**Table S1.** Parameters of DACs used in high-pressure experiments.

| DAC number | DAC's diameter, material of the DAC | Culet diameter / Heated at pressure | Initial sample composition, insulating gasket | Experiments |
|---|---|---|---|---|
| Y1 (pulsed field cell) | d = 15 mm, NiCrAl | 50 um / 195 GPa | Y/AB, $CaF_2$/epoxy/W | Pulsed field measurements Transport measurements |
| Y2 (electrical cell) | d = 25 mm, BeCu | 50 um / 205→213 GPa | Y/AB, $CaF_2$/epoxy/W | Transport measurements |
| Y3 (RF cell) | d = 39 mm, BeCu | 75 um / 180→165 GPa | LiAB/$YCl_3$/$Y(AB)_3$, $TaW_{10\%}$/$Ta_2O_5$ | Radio-frequency measurements X-ray diffraction (SPring8, 2026) |
| Y4 (RF cell) | d = 39 mm, BeCu | 75 um / 148→140 GPa | Y/AB, $TaW_{10\%}$/$Ta_2O_5$ | Radio-frequency measurements |
| Y5 (electrical cell) | d = 25 mm, BeCu | 50 um / 174 GPa | $Y_3Pd$ /AB, $CaF_2$/epoxy/W | Transport measurements |
| Y6 (RF + electrotransport cell) | d = 39 mm, NiCrAl | 30 um / 180 GPa | Y/Pd(film)/AB, $TaW_{10\%}$/$Ta_2O_5$ | Radio-frequency measurements Transport measurements X-ray diffraction (SPring8, 2026) |
| Y7 (electrical cell) | d = 25 mm, BeCu | 50 um / 174 GPa | YAl/AB, $CaF_2$/epoxy/W | Transport measurements |

*Theoretical calculations*

All electronic-structure calculations were carried out within density functional theory (DFT) [19, 20], as implemented in the Quantum ESPRESSO (QE) package [12,13]. The electron–ion interaction was described using scalar relativistic optimized norm-conserving Vanderbilt (ONCV) pseudopotentials[14,15] from the PseudoDojo database[16]. Exchange–correlation effects were treated at the generalized gradient-approximation (GGA) level using the Perdew–Burke–Ernzerhof (PBE) functional[17]. The kinetic-energy cutoff for the plane-wave basis was set to 80 Ry for $YH_6$ and to 90 Ry for $YH_9$ and $YH_{10}$. Self-consistent-field (SCF) calculations were converged to an energy threshold of $10^{-10}$ Ry. The Brillouin zone was sampled using Γ-centered Monkhorst–Pack k-point meshes[18]. We employed a 16×16×16 grid for $YH_6$ and $YH_{10}$, and a 24×24×16 grid for $YH_9$. Metallic occupations were treated with Methfessel–Paxton smearing[19], using a smearing width of 0.04 Ry for $YH_6$ and 0.02 Ry for $YH_9$ and $YH_{10}$. Smearing was selected for each compound individually based on convergence tests (see, for example, Figure S2). With these parameters, the total energies were converged to better than 1 meV/atom. Structural relaxations were performed for both lattice parameters and internal atomic coordinates until the changes in total energy and the residual forces were below $10^{-7}$ Ry and $10^{-6}$ Ry/$a_0$, respectively.

Phonon dispersions and el-ph coupling matrix elements were calculated within density functional perturbation theory (DFPT) [20]. The self-consistency threshold used in the phonon calculations was set to $10^{-16}$ Ry. The dynamical matrices and electron–phonon matrix elements were evaluated on homogeneous q-point grids of 6×6

×6 for $YH_6$ and $YH_9$, and 8×8×8 for $YH_{10}$. To improve the convergence of the superconducting properties with respect to the electronic smearing and Brillouin zone sampling, we rescaled the Eliashberg spectral function following the procedure described in Ref. [21]. The corrected spectral function is defined as

$$\alpha^2 F_r(\omega) = \frac{N_F^{tetra}}{N_F^{smear}} \alpha^2 F(\omega). \qquad (S1)$$

In this expression, $N_F^{tetra}$ denotes the density of states at the Fermi level obtained from a well-converged tetrahedron method calculation on a dense k-point mesh, whereas $N_F^{smear}$ is the corresponding value extracted using the Gaussian smearing adopted in the el-ph calculations. The quantity $\alpha^2F(\omega)$ therefore represents the rescaled Eliashberg spectral function.

The total el-ph coupling constant was then computed by integrating the rescaled spectral function over the full phonon spectrum,

$$\lambda = 2 \int \frac{\alpha^2 F_r(\omega)}{\omega} d\omega. \qquad (S2)$$

For the final integration of the el-ph matrix elements, dense electronic meshes of 48×48×48 were used for $YH_6$ and $YH_{10}$, while a 48×48×32 grid was adopted for $YH_9$. The double delta integrations were performed using 50 smearing values ranging from 0.001 to 0.05 Ry (example is given in Figure S2).

The superconducting properties were obtained by solving the isotropic Migdal–Eliashberg equations within the full bandwidth approximation [22], using the IsoME code [23]. The Matsubara frequency cutoff was fixed at 7 eV which is much higher than the maximum phonon frequency in the hydrides under consideration. In calculations including the energy dependence of the electronic density of states, the chemical potential was updated self consistently. An energy window of 5 eV was used for all quantities, except for the energy shift and the chemical potential update, for which a reduced cutoff of 2 eV was employed. In both cases, these intervals are more than an order of magnitude larger than the expected values of the superconducting gap $\Delta(0)$.

The superconducting density functional theory (SCDFT) calculations were done with the SPG2020 functional for the phononic parts [24] and the Coulombic part with the plasmonic correction [25] as implemented with Eq.(3) of Ref. [26]. The detailed calculation settings are summarized in Table S2.

**Table S2.** Detailed conditions for calculating $T_c$'s of yttrium hydrides within the SCDFT approach.

| | | **$YH_6$ (200 GPa)** | **$YH_9$ (200 GPa)** | **$YH_{10}$ (250 GPa)** |
|---|---|---|---|---|
| Unitcell | | Primitive (BCC) | Conventional (hexagonal) | Primitive (FCC) |
| charge density | k | 16x16x16 equal mesh | 24x24x16 equal mesh | 16x16x16 equal mesh |
| | interpolation | 1st order Hermite Gaussian [19] with width 0.040 Ry | 1st order Hermite Gaussian with width 0.020 Ry | 1st order Hermite Gaussian with width 0.020 Ry |
| dielectric matrix $\varepsilon_{ij}$ | k for bands crossing $E_F$ | 21x21x21 equal mesh | 15x15x9 equal mesh | 15x15x15 equal mesh |
| | k for other bands | 7x7x7 equal mesh | 5x5x3 equal mesh | 5x5x5 equal mesh |
| | number of unoccupied bands (within $E_F$ + 70 eV) | 52 | 121 | 61 |
| | interpolation | Tetrahedron with the Rath-Freeman treatment [27] | | |
| DOS for phononic kernels | K | 23x23x23 equal mesh | 19x19x13 equal mesh | 19x19x19 equal mesh |
| | interpolation | Tetrahedron with the Blöchl correction [28] | | |
| SCDFT gap function | number of unoccupied bands (within $E_F$ + 70eV) | 52 | 121 | 61 |
| | k for the electronic kernel | 7x7x7 equal mesh | 5x5x3 equal mesh | 5x5x5 equal mesh |
| | k for the KS energies | 23x23x23 equal mesh | 19x19x13 equal mesh | 19x19x19 equal mesh |
| | Sampling points [29] for bands crossing $E_F$ | 6000 | | |
| | Sampling points for the other bands | 200 | | |
| | Sampling error in $T_c$ (%) | ~2.8 | ~3.1 | ~2.4 |

**Table S3.** Comparison of SCDFT calculation results with two functionals: LM2005 and SPG2020. Note that the current SCDFT algorithm involves random k-point sampling and yields a few % of statistical error.

| | $YH_6$ (200 GPa) | $YH_9$ (200 GPa) | $YH_{10}$ (250 GPa) |
|---|---|---|---|
| LM2005 | 152 | 185 | 214 |
| SPG2020 | 175 | 191 | 221 |
| Experiment | 218-226 | 236-243 | < 260-270 |

Anharmonic vibrational properties were computed using the stochastic self-consistent harmonic approximation (SSCHA) [30]. This variational method treats quantum and thermal fluctuations of the nuclei and anharmonicity of the Born–Oppenheimer potential energy surface non-perturbatively, while remaining within the Born–Oppenheimer approximation. The ionic free energy is minimized with respect to centroid positions and an auxiliary force-constant matrix, and anharmonic phonons are obtained from the free-energy Hessian within the bubble approximation.

We performed SSCHA minimizations in fixed-cell (constant-volume) mode. The ionic centroids and auxiliary force-constant matrix were optimized at fixed lattice vectors. Energies, forces, and stresses for each stochastic configuration were evaluated with a purpose-trained MTP [31]. Using this type of MLIP in place of direct DFT force evaluations reduced the computational cost of the minimization by several orders of magnitude.

Supercells were constructed from the relaxed primitive structures, using 2×2×2, 3×3×3, and 4×4×4 expansions for $YH_6$, $YH_9$, and $YH_{10}$. Each minimization was initialized with a stochastic population of 1000–3000 configurations, which was gradually increased to 10,000–15,000 configurations in the final population depending on the material and supercell size. The final population size was chosen such that the maximum change in anharmonic phonon energies between the last two populations was below 1 meV.

Resampling of the ensemble was triggered when the effective sample size $n_{eff}$ fell below 0.6. Gradient updates were performed using the ROOT2 algorithm with preconditioning. Convergence was controlled through the meaningful-factor parameter, set to $10^{-7}$–$10^{-8}$, which defines convergence when the ratio between the free-energy gradient and the auxiliary dynamical matrices falls below this threshold.

The MTP used for $YH_6$ was trained on a DFT dataset containing 250 configurations generated from a 2×2×2 supercell at 200 GPa. For $YH_9$, the potential was trained on 400 configurations from a 2×2×2 supercell at 200 GPa, with the training procedure being roughly one order of magnitude more computationally demanding than for $YH_6$. For $YH_{10}$, the MTP was trained on a combined dataset consisting of 250 configurations from a 2× 2× 2 supercell and 100 configurations from a 3×3×3 supercell, both generated at 250 GPa.

We previously developed a theory for describing the instability of phonon mediated superconductors, particularly polyhydrides, with respect to strong electron-phonon interaction. In this framework, the metallic state becomes unstable toward a reconstruction of the hydrogen sublattice, which may lower the crystal symmetry, lead to partial hydrogen loss, and open a gap or a pronounced depression in the electronic density of states near the Fermi level, replacing the superconducting gap [32-34]. The key quantity is the stability parameter ξ,

$$\xi \equiv \max_{T}\left\{\int_0^{\infty} g\left(\frac{\omega}{2\pi T}\right)\frac{2\alpha^2F(\omega)}{\omega}d\omega\right\} < 1, \qquad (S3)$$

where $g(x) = 6x^2 + 12x^3 Im\,\psi'(ix) + 6x^4 Re\,\psi''(ix)$, and $\psi(x)$ is the digamma function [32]. The condition $\xi < 1$ is the absolute stability condition: at $\xi = 1$ the electronic specific heat vanishes, and for $\xi > 1$ the metallic state is unstable with respect to any deviations from thermal equilibrium. The actual phase transition between two states with the higher and lower Gibbs free energy, is expected to occur earlier, at a critical value $\xi_c < 1$, because it is a first-order transition with a kinetic barrier, which requires a finite time for a passing through.

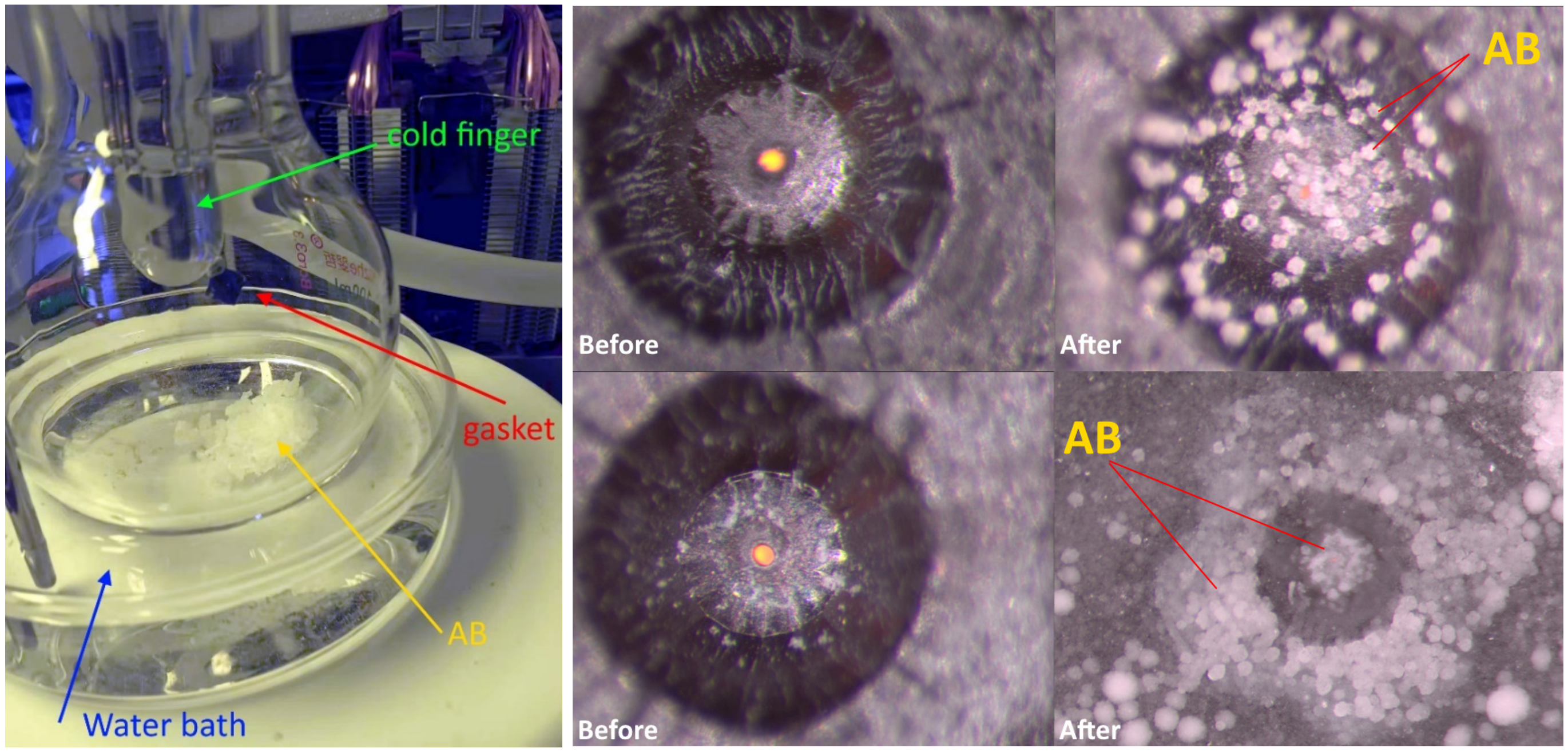

**Figure S1.** (left) Sublimation of $NH_3BH_3$ using a system containing a water bath, a heated vacuum flask with a water-cooled cold finger. (right) Sublimating $NH_3BH_3$ directly into the gasket hole was used in this study. The idea was to automatically load 4-5 or more gaskets with fresh ammonia borane without manual intervention. The gaskets were glued to the cold finger of the sublimation system. Sublimation was carried out at 50-60°C for 4 hours. As a result, the gaskets were filled with a layer of fairly large (10-30 μm) ammonia borane crystals.

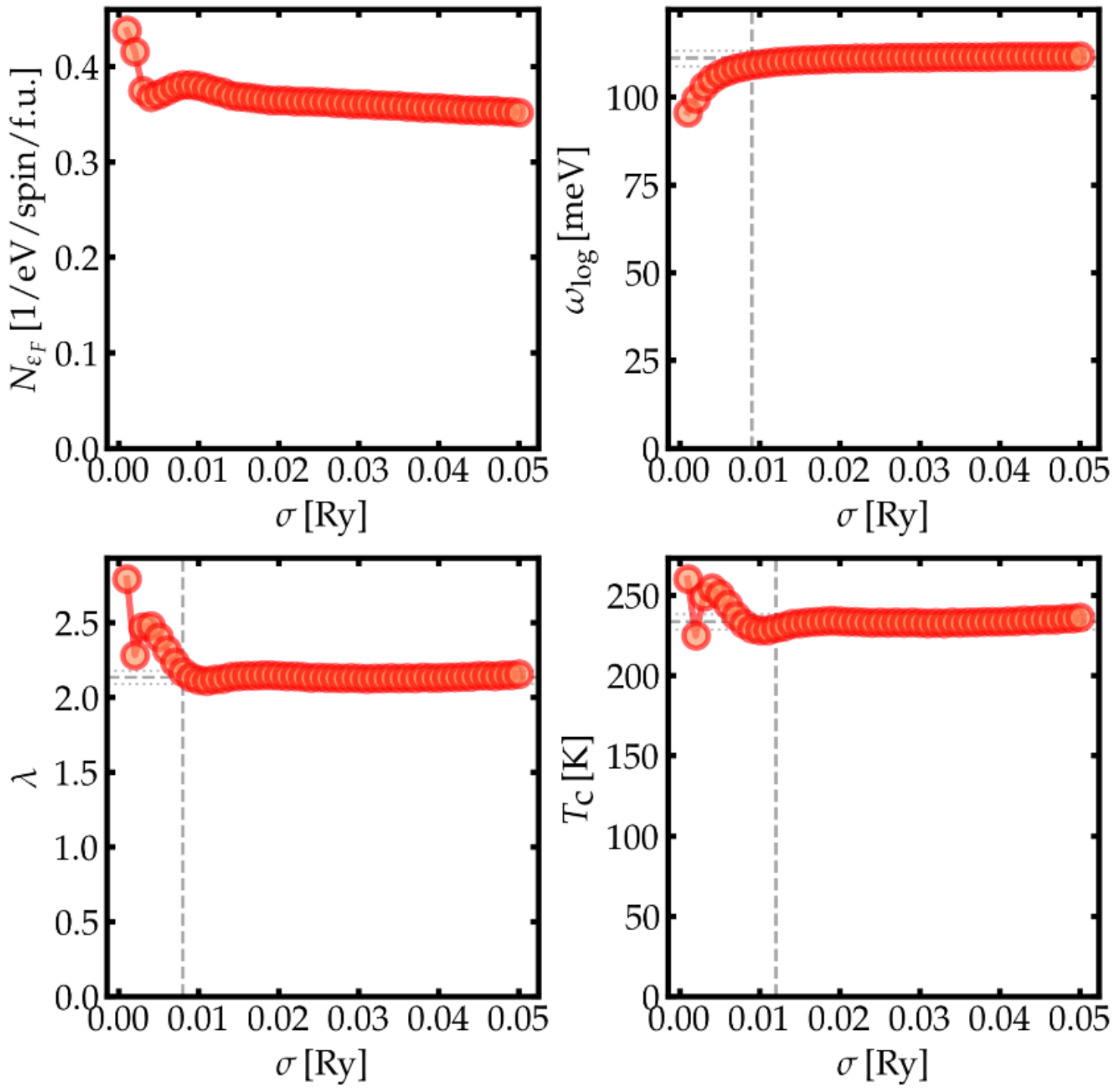

**Figure S2.** Example of convergence tests for the renormalized electron-phonon interaction parameters of cubic $YH_{10}$ at 250 GPa including SSCHA corrections. Dependence on sigma broadening of various electron-phonon interaction parameters, density of states and Allen-Dynes critical temperature for 8×8×8 q-mesh.

## II. Supplementary transport data

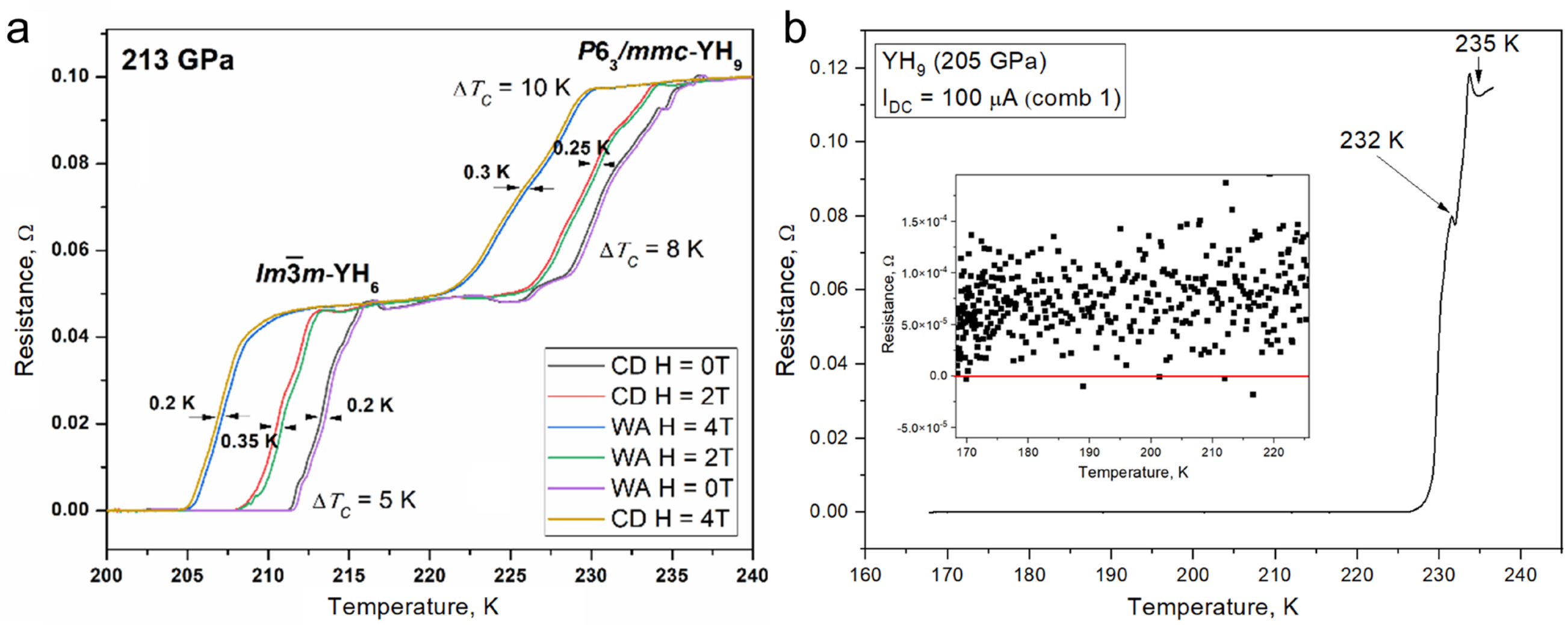

**Figure S3.** Temperature dependence of electrical resistance and superconducting transitions in yttrium hydrides under high pressure in the DAC Y2. (a) Temperature-dependent electrical resistance of a mixed-phase yttrium hydride sample in DAC Y2 at a pressure of 213 GPa. The curve reveals two distinct superconducting steps corresponding to the cubic *bcc*-$YH_6$ and the hexagonal $YH_9$ phase. Data are shown for cooling down (CD) and warming up (WA) pathways under various external magnetic fields ($B$ = 0, 2, and 4 T). The figure shows almost zero thermal hysteresis (0.2-0.35 K) in superconducting transitions in hydrides and their good reproducibility in warming and cooling cycles. The annotations also highlight the SC transition widths ($\Delta T$). (b) Temperature dependence of the electrical resistance for the $YH_9$ phase at 205 GPa, recorded using a direct current $I_{DC}$ = 100 μA (contact configuration 1). The figure shows that along with *hcp*-$YH_9$ with $T_c$ = 235 K, in the Y-H system above 200 GPa, there may be another phase with a lower $T_c$ = 232 K, still different from $YH_6$. Inset: magnified view of the resistance baseline from 170 K to 225 K, demonstrating a state of near-zero resistance relative to the zero-level reference (red line) within the experimental noise limits.

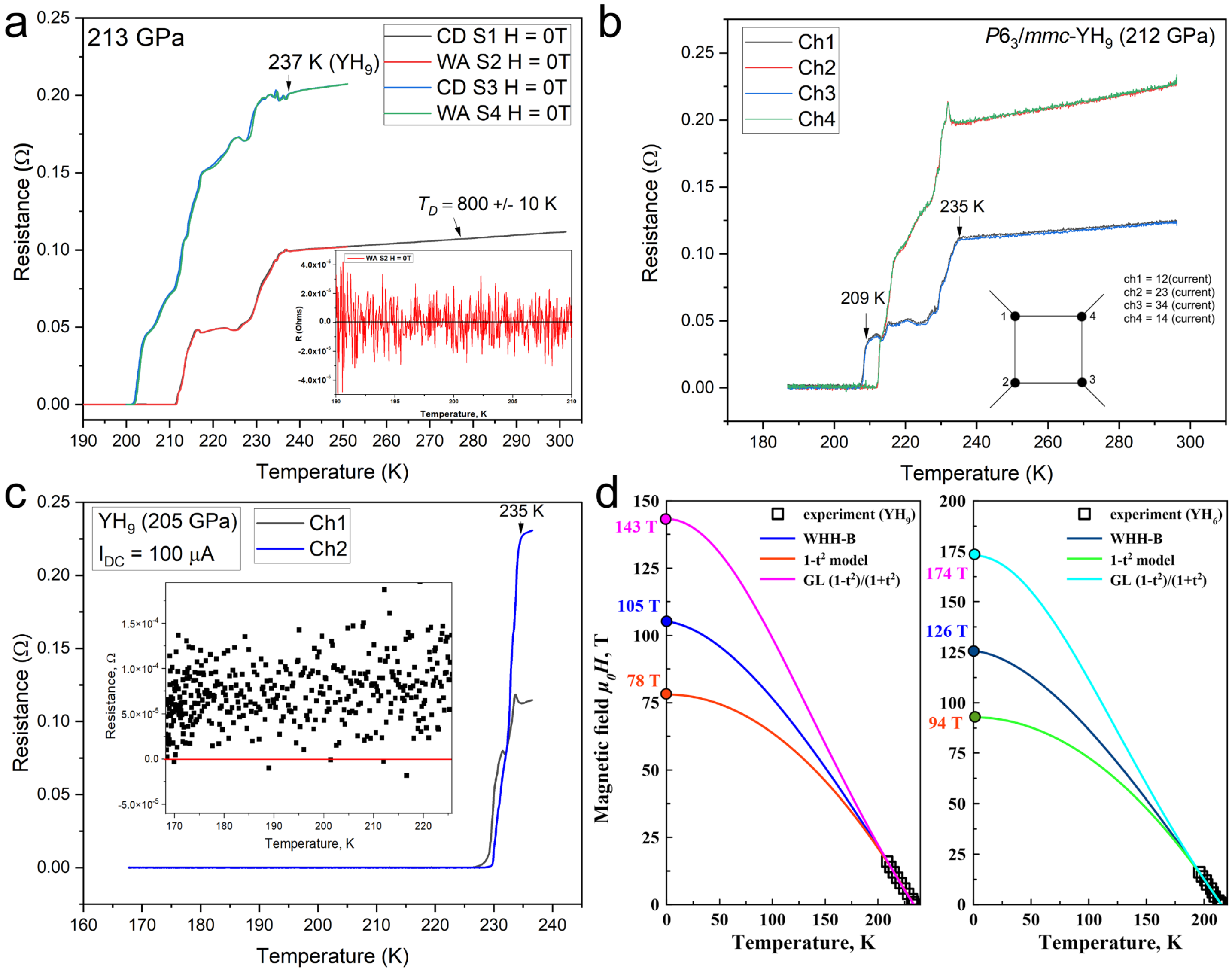

**Figure S4.** Evolution of the results of multichannel resistivity measurements in the DAC Y2 sample depending on the pressure in the range from 213 GPa to 205 GPa. Additional laser heating was performed to improve its homogeneity, which, however, did not lead to significant progress. This indicates that the additional steps in *R(T)* are due to thermodynamically stable phases, which are more complex in their structure than $YH_9$ or $YH_6$. (a) Two-channel measurements of the temperature dependence of resistance at 213 GPa in heating and cooling cycles with minimal thermal hysteresis in zero magnetic field. Inset: residual resistance below 210 K. Onset $T_c$($YH_9$) determined from the onset of the anomalous zone of *R(T)* is 237 K, in good agreement with the results of the previous work Ref. [35]. (b) Four-channel measurements at 212 GPa after additional laser heating. Inset: current and voltage connection diagram. The sample remains inhomogeneous. It is important to note that the different channels provide additional nontrivial information about the sample, not even coinciding with respect to $T_c$(onset) and $T_c$(offset). (c) Two-channel measurements at 205 GPa after additional laser heating. The sample composition is significantly simplified, with only one additional step present at 232 K. (d) Temperature dependence of the upper critical field of the components of the DAC Y2 sample at 213 GPa, determined from Tc(onset) together with extrapolation of experimental data using various theoretical models (GL – Ginzburg-Landau).

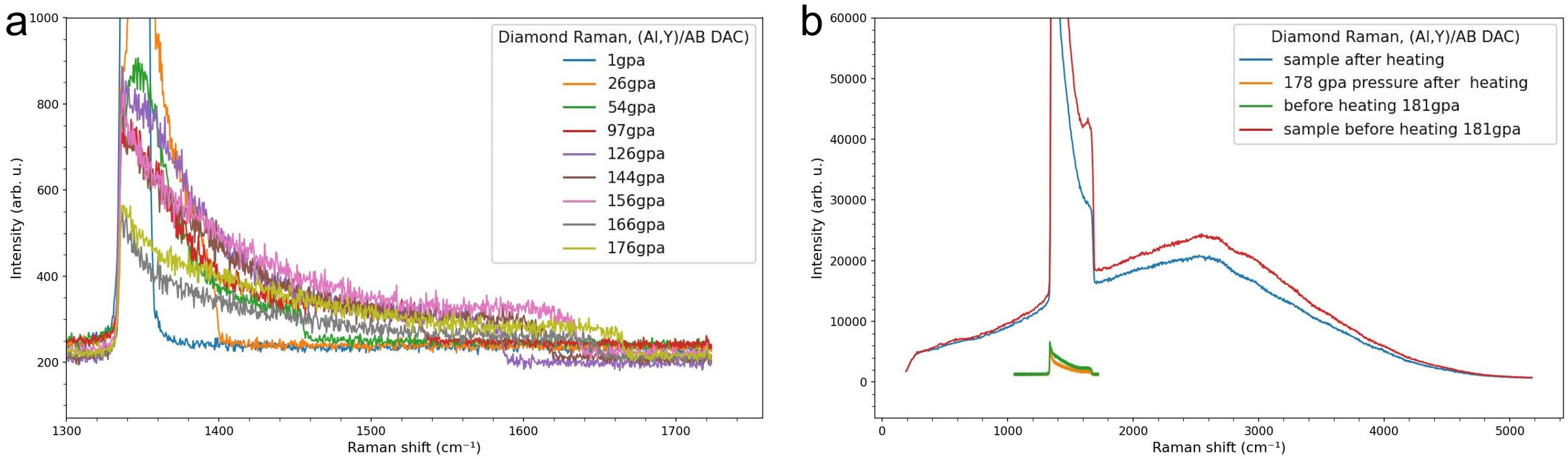

**Figure S5.** Raman measurements of a DAC Y7 with an yttrium-aluminum alloy during compression: (a) before laser heating), and (b) after laser heating. The pressure after laser heating was approximately 178 GPa. After laser heating, the intensity of the Raman band of N-H and B-H vibrations in amorphous ammonia borane decreases significantly. The absence of a hydrogen signal is typical for synthesis near 2 Mbar on 50-μm diamond anvil culets: all the hydrogen usually reacts completely with the metal particle.

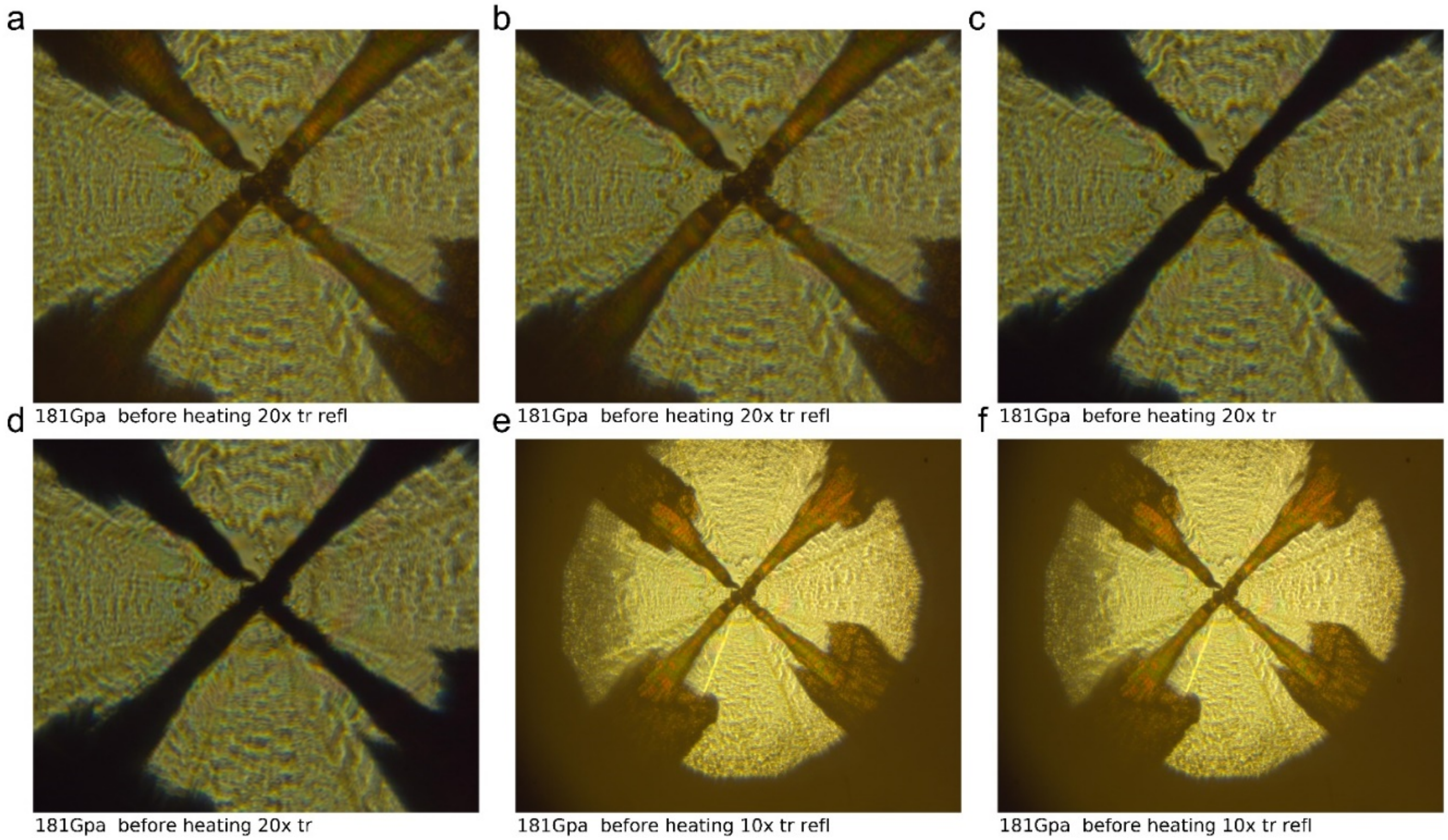

**Figure S6**. Optical photos of DAC Y7's high-pressure chamber with AlY alloy (may contain Y-Al intermetallics) loaded with $NH_3BH_3$, and Ta/Au electrodes connected to the sample before laser heating in reflected ("refl") and transmitted ("tr") light.

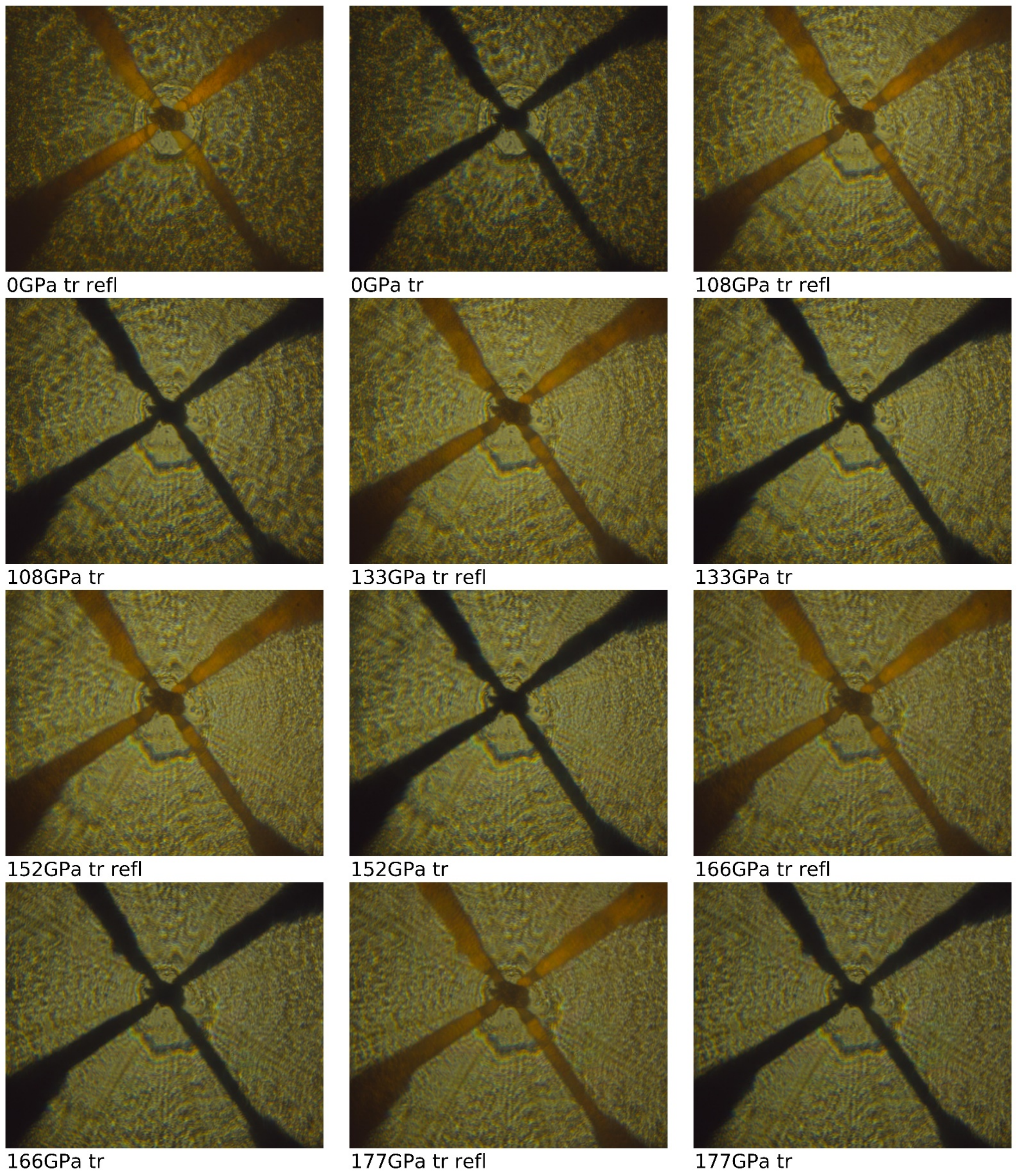

**Figure S7**. A series of optical photographs of the $Y_3Pd$/AB sample and the DAC Y5 electrode system before laser heating at different pressures. ("tr" – transmitted light, "refl" – reflected light).

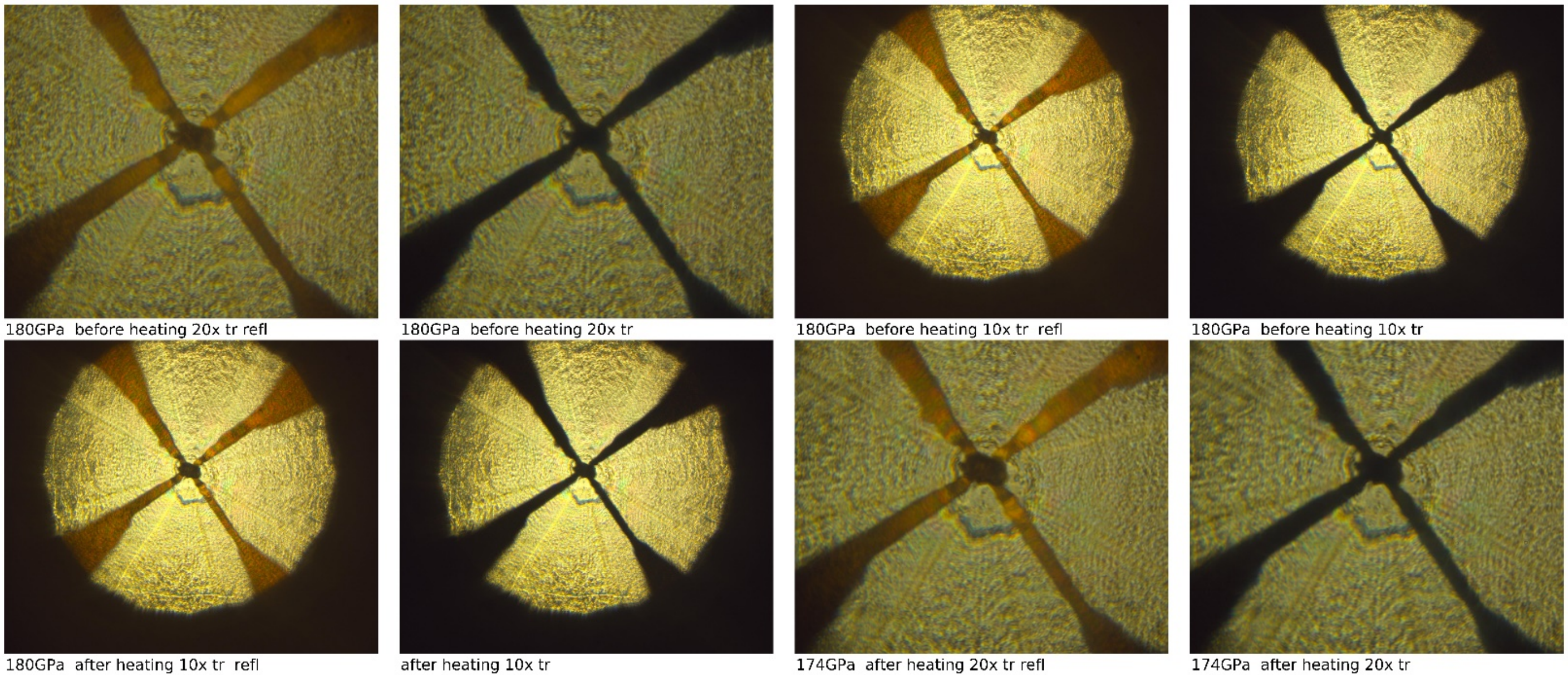

**Figure S8**. A series of optical photographs of the $Y_3Pd$/AB sample and the DAC Y5 electrode system before laser heating at 180 GPa and after the laser heating ("tr" – transmitted light, "refl" – reflected light).

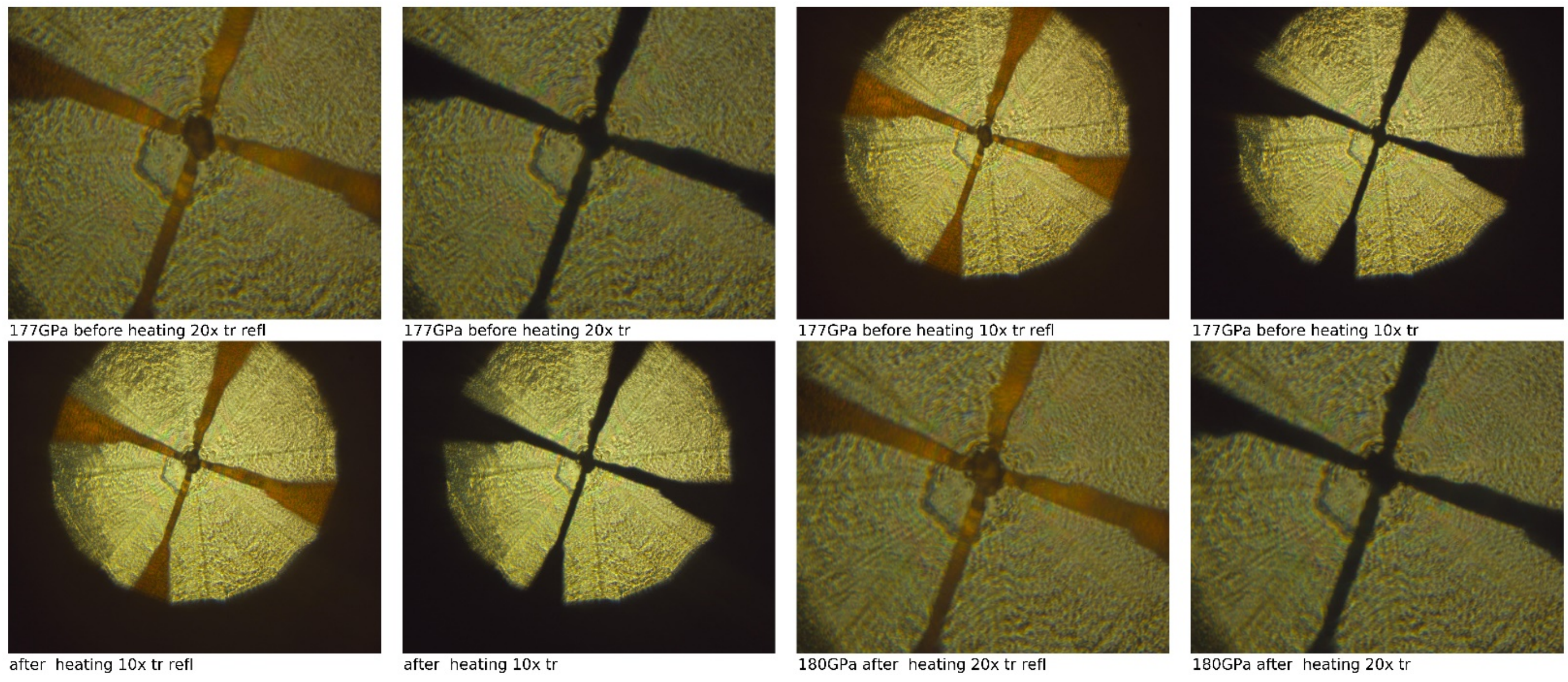

**Figure S9**. A series of optical photographs of the $Y_3Pd$/AB sample and the DAC Y5 electrode system before laser heating at 177 GPa and after the laser heating ("tr" – transmitted light, "refl" – reflected light).

Despite numerous attempts to laser heat (Y,Al) and $Y_3Pd$ samples in DACs Y5, Y7, high-temperature superconductivity has not yet appeared in them.

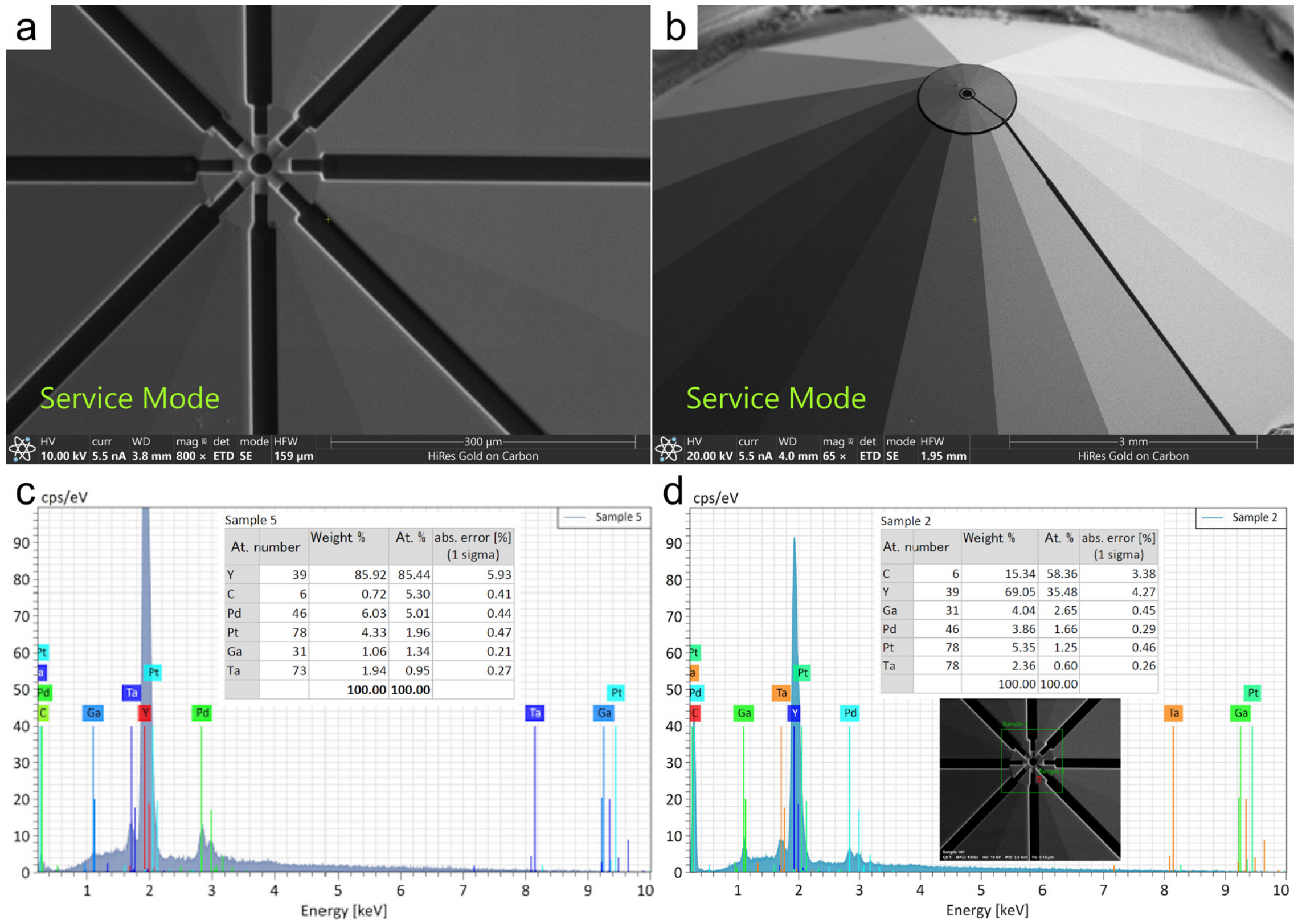

Sample 5

| | At. number | Weight % | At. % | abs. error [%] (1 sigma) |
|---|---|---|---|---|
| Y | 39 | 85.92 | 85.44 | 5.93 |
| C | 6 | 0.72 | 5.30 | 0.41 |
| Pd | 46 | 6.03 | 5.01 | 0.44 |
| Pt | 78 | 4.33 | 1.96 | 0.47 |
| Ga | 31 | 1.06 | 1.34 | 0.21 |
| Ta | 73 | 1.94 | 0.95 | 0.27 |
| | | **100.00** | **100.00** | |

Sample 2

| | At. number | Weight % | At. % | abs. error [%] (1 sigma) |
|---|---|---|---|---|
| C | 6 | 15.34 | 58.36 | 3.38 |
| Y | 39 | 69.05 | 35.48 | 4.27 |
| Ga | 31 | 4.04 | 2.65 | 0.45 |
| Pd | 46 | 3.86 | 1.66 | 0.29 |
| Pt | 78 | 5.35 | 1.25 | 0.46 |
| Ta | 78 | 2.36 | 0.60 | 0.26 |
| | | 100.00 | 100.00 | |

**Figure S9.** Scanning electron microscopy of Y/Pd(film) sample, electrode system and Lenz lens of DAC Y6. (a) Electron micrograph of 30 um culet with 8 electrodes and Y/Pd ring shaped by Ga FIB. In the first step, we sputtered Ta/Pt electrodes, then yttrium (about 0.5-1 um) was deposited, which was then covered with 50-100 nm layer of Pd. In the final step, a central hole was cut out, and shorting bridges between the electrodes were removed by Ga FIB. (b) Electron micrograph of Lenz lens formed on the second diamond anvil. (c) Elemental EDX analysis of Ta/Pt/Y/Pd electrode near the culet. Inset: percentage abundance of different elements. (d) Elemental analysis of the whole sample and electrodes within 150 μm radius from the culet center to determine the film thickness of sputtered palladium.

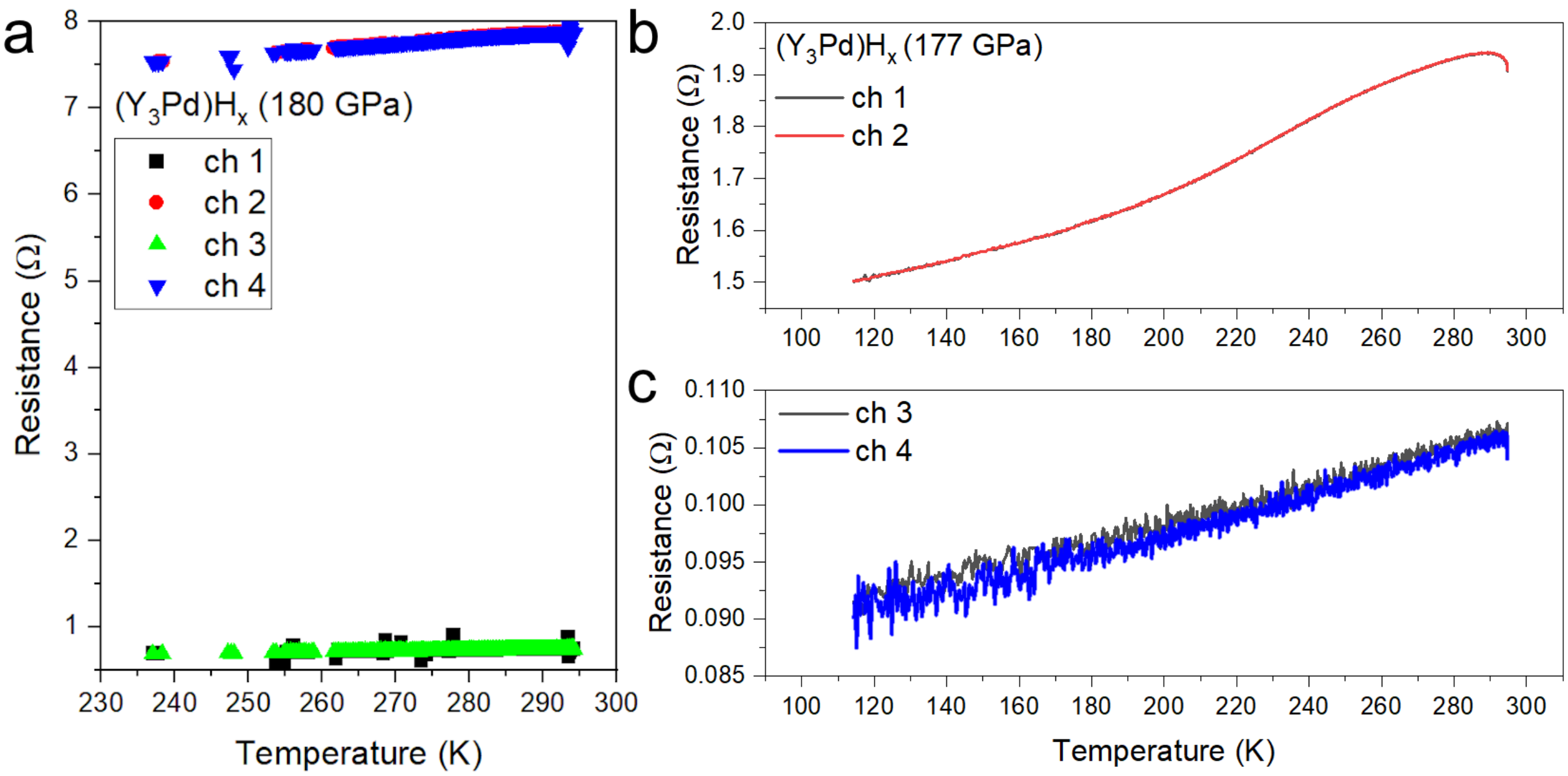

**Figure S10.** Multi-channel measurements of the temperature dependence of the electrical resistance of the sample $(Y_3Pd)H_x$ (DAC Y5) at DC current (0.1 mA, delta mode): (a) at 180 GPa; there is a certain instability of the electrode system. (b, c) At 177 GPa. There are no signs of high-temperature superconductivity.

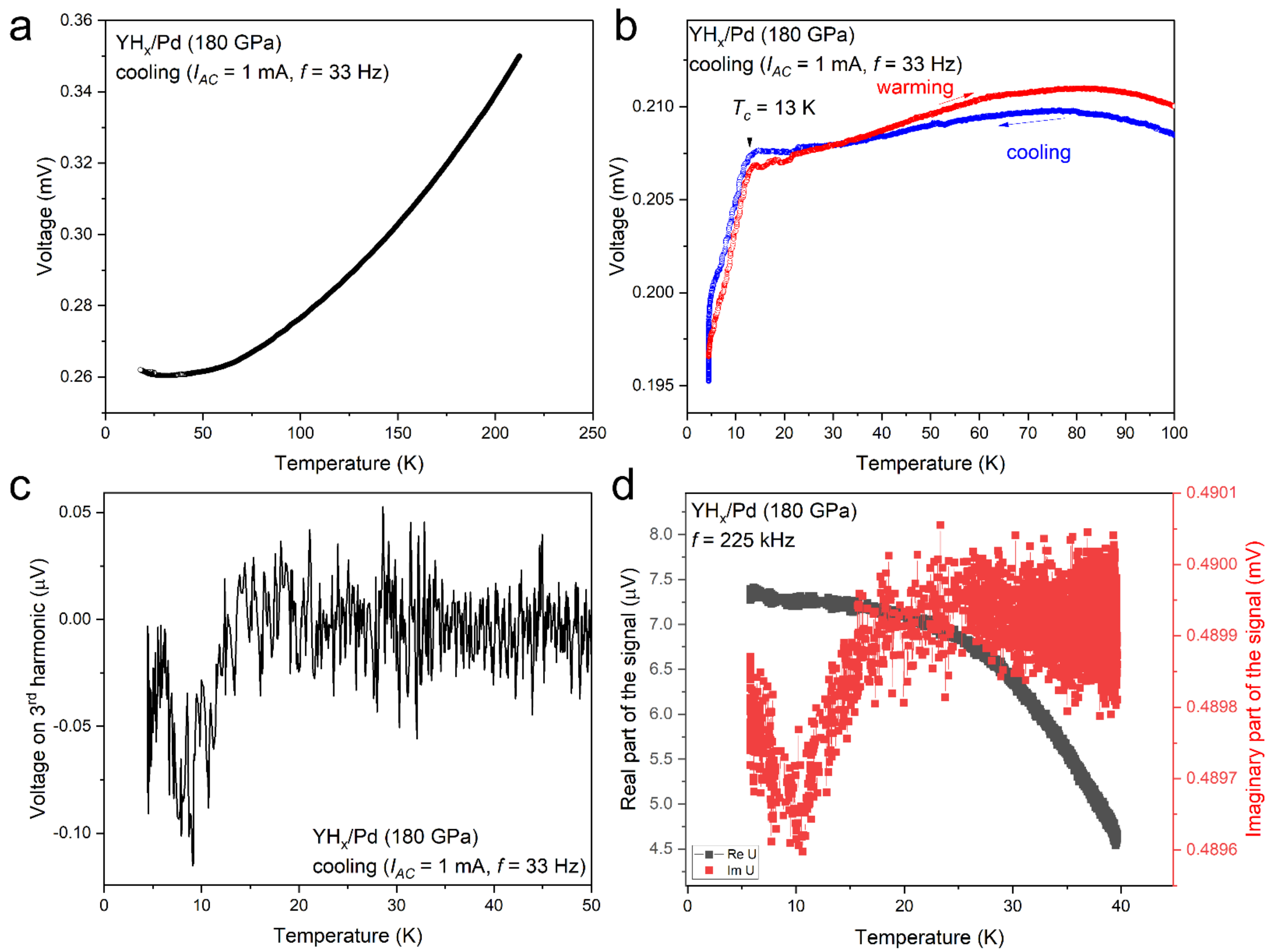

**Figure S11.** Study of the electrical properties of an yttrium hydride sample DAC Y6 coated with a nanometer-sized palladium film (Y/Pd precursor) at 180 GPa after repeated intense laser heating. (a) Electrical resistance over a wide temperature range. (b) A sharp drop in the electrical resistance of the sample below 13 K during cooling and heating cycles. The study was performed with an alternating current of 1 mA and a frequency of 33 Hz. (c) Simultaneously, the emergence of a signal at the third harmonic (3×33 = 99 Hz) is observed, a nonlinear effect that often accompanies the appearance of superconductivity. (d) Radio-frequency study of the same sample, performed through built-in Lenz lens at a frequency of 225 kHz.

The third harmonic occurs when the sample is heated by a passing current. Indeed, if the current source produces a sinusoidal current $I = I_0 \sin(\omega t)$, then the heating of the sample with resistance $R$ will be determined by the power $P(t) = I_0^2 R(\sin(\omega t))^2$, which causes periodic temperature changes and modulation of the electrical resistance of the sample: $R = R_0(1 + aP(t)dR/dT)$, where we have added an oscillating part to the resistance *R(T)*. Since *dR/dT* becomes large precisely in the region of the superconducting transition, a significant third harmonic is observed in the detected voltage drop $V = RI = R_0 I_0 \sin(\omega t) + aR_0 I_0 (dR/dT)(\sin(\omega t))^3$.

We also attempted to detect the Seebeck effect in the $YH_6$ sample (DAC Y1) by passing a strong AC current $I_{AC}$ = 10 mA through a pair of contacts and detecting the resulting thermoelectric EMF (Figure S12). The Seebeck effect is the emergence of electromotive force (EMF) that develops across two points of an electrically conducting material when there is a temperature difference between them. AC current provokes temperature gradient oscillations $grad\, T \propto I_0^2 R(\sin(\omega t))^2 = 0.5 I_0^2 R(1 - \cos(2\omega t))$, which can be detected at the second harmonic of the heating current frequency. Since the Seebeck effect is linearly proportional to the temperature gradient, $V \propto S\ grad\ T$, the detected signal will also have a component ($V_{2\omega}$) at a frequency of $2\omega$ proportional to $S$.

The problem in detecting the Seebeck effect in high-pressure diamond anvil cells is that only a small component of the 2-contact resistance $R = R_{wire} + R_{hydride}$ is due to the hydride sample (typically ~ 1%). Therefore, when measuring the second harmonic ($V_{2\omega}$), only a small fracture (due to a change in $S = S_{\mathrm{wire}} + S_{\mathrm{hydride}}$) or a small step (due to a change in $R$) is observed in the experiment at the temperature of the superconducting transition (Figure S12).

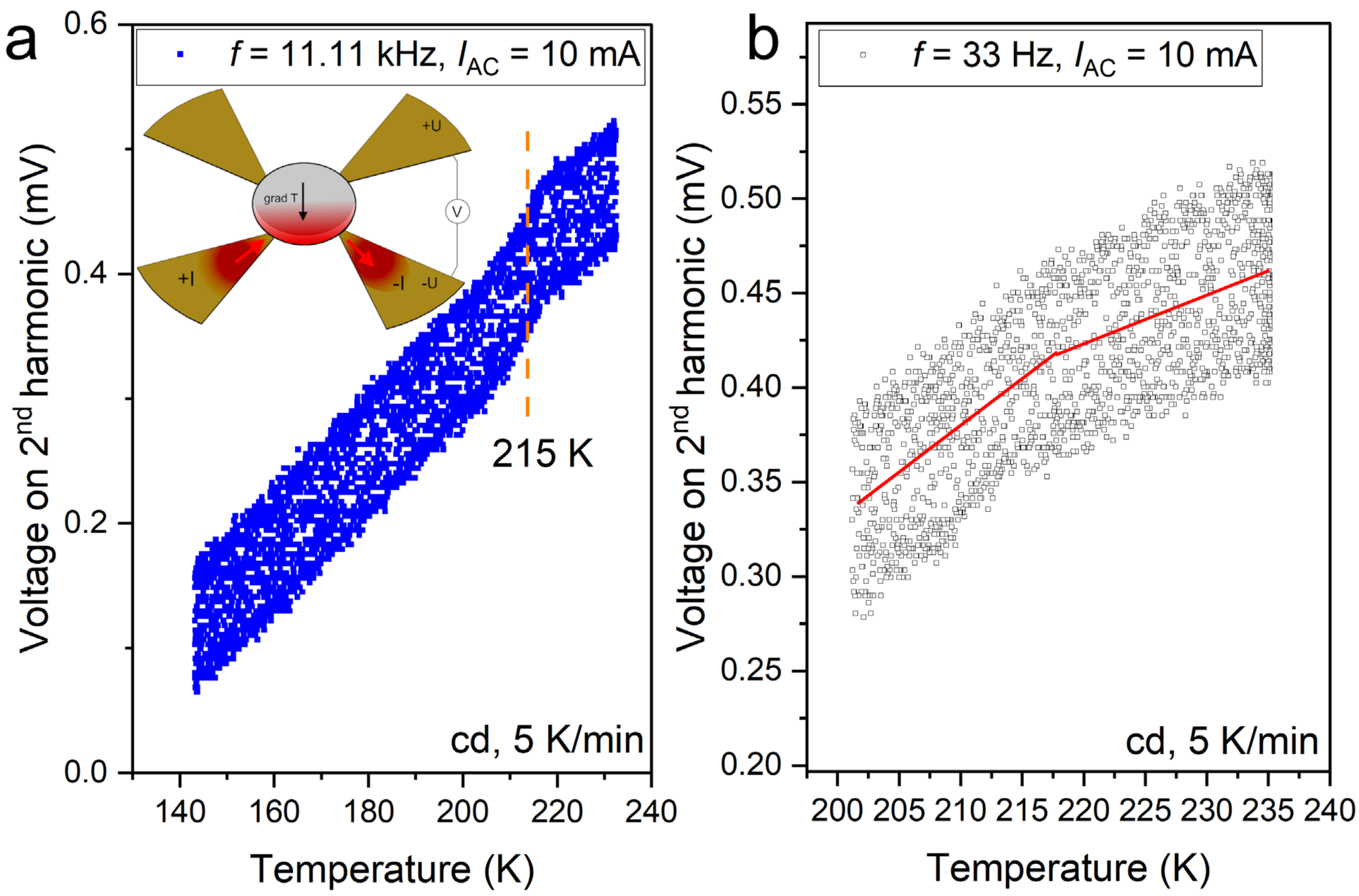

**Figure S12.** Detection of the Seebeck effect in $YH_6$ sample in DAC Y1. (a) Temperature dependence of the second-harmonic voltage response measured at a high modulation frequency of 11.11 kHz with an applied alternating current $I_{AC}$ = 10 mA during a cooling cycle (cd at 5 K/min). The dashed orange line indicates a small step near 215 K, which corresponds to the superconducting transition in $YH_6$. Inset: schematic diagram of the experimental contact configuration used for the second-harmonic thermoelectric measurements. An alternating current passed through the current leads (+I, -I) induces localized periodic Joule heating, creating a temperature gradient (grad $T$) across the sample, while the resulting Seebeck voltage is detected across the voltage contacts (+U, -U). (b) Temperature dependence of the second-harmonic voltage signal recorded at a lower frequency of 33 Hz ($I_{AC}$ =10 mA, cooling rate is 5 K/min) near the superconducting transition region. The red curves represent linear fit of the data below and above the transition temperature.

We also attempted to measure the differential conductivity for a sample $YH_6$ at ≈200 GPa using a single, particularly thin and resistive electrode. It was hypothesized that such an electrode might exhibit point-contact properties and serve to measure the superconducting gap $\Delta(T)$. However, it turned out that the shape of the differential conductivity peak changes irregularly as the sample cools, making it impossible to extract data on the superconducting gap (Figure S13).

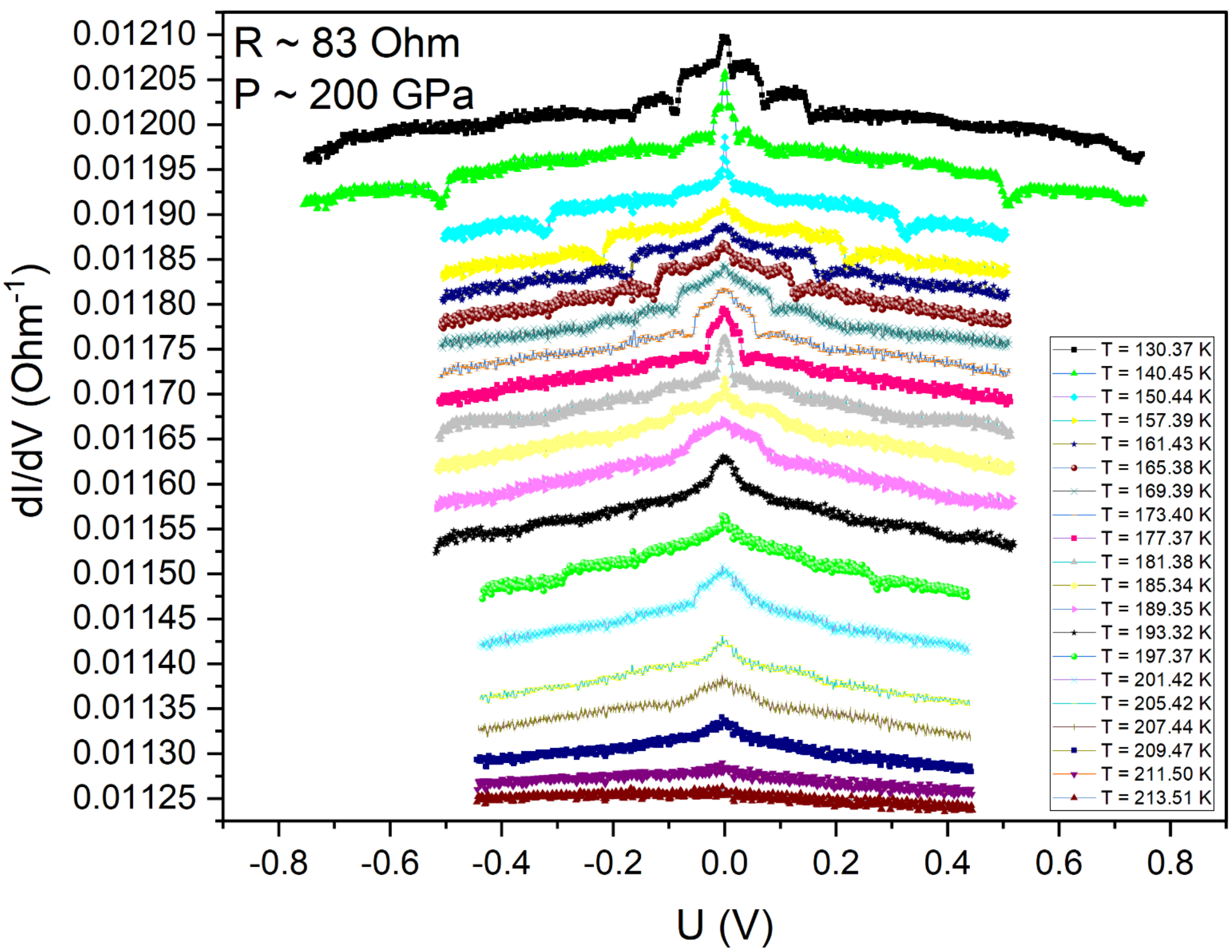

**Figure S13.** An attempt to perform point contact spectroscopy on a $YH_6$+$YH_9$ (DAC Y2) sample at 200 GPa. Below 213.5 K, a state of "zero" (below the thermal noise level) residual resistance is achieved in the sample. Applying a DC bias voltage to the high-resistance contact (83 Ω) also causes a change in the *V(I)* current-voltage characteristic of the point contact relative to a small periodic AC signal passing through the same contact. However, in this case, it turned out that the high-resistance contact is not ballistic, and instead of the superconducting gap parameters, we obtain a value proportional to the critical contact current $I_c$, which increases with decreasing temperature.

## III. Supplementary radio-frequency data

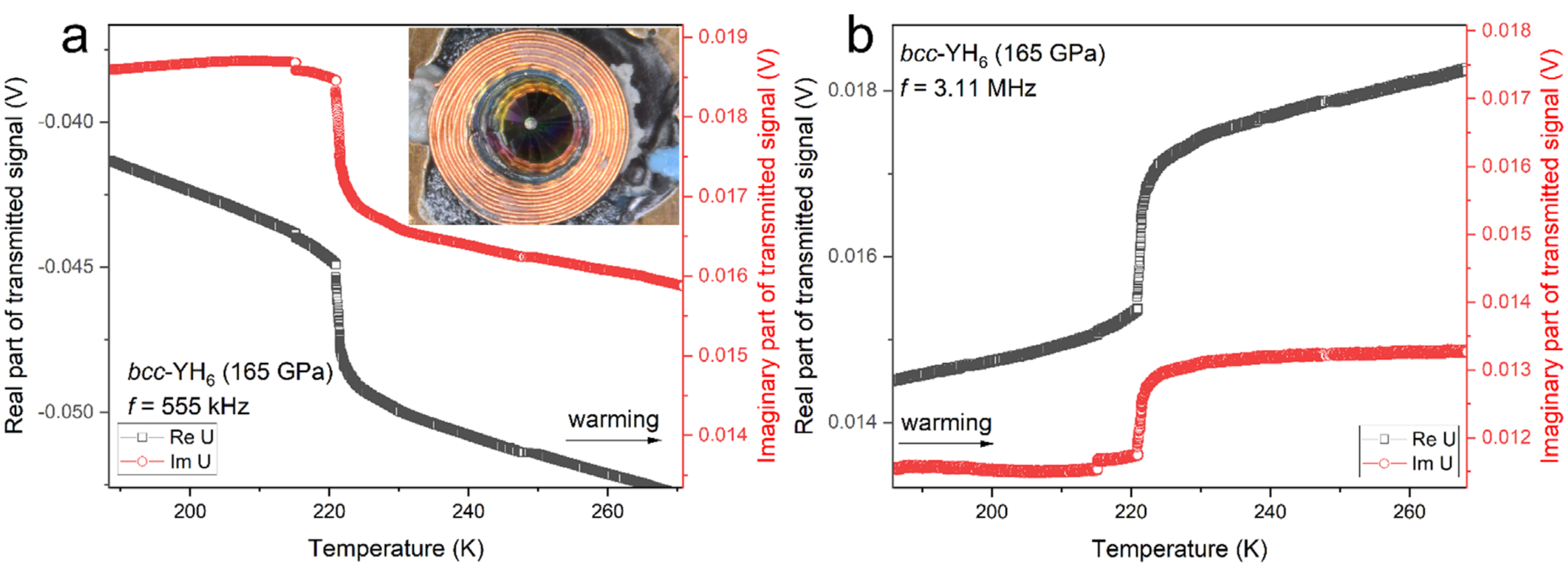

**Figure S14.** Radio-frequency transmission measurements of *bcc*-$YH_6$ under pressure of 165 GPa (DAC Y3). (a) Temperature dependence of the real part (Re U, black open squares) and imaginary part (Im U, red open circles) of the transmitted RF signal measured at a frequency of $f = 555$ kHz. The distinct steps near 221 K appeared during the warming cycle indicate the superconducting transition. Inset: microphotograph of the sensitive multiturn coil ($N \approx$ 20-30) glued to a diamond anvil with Lenz lens. (b) Transmitted RF signal (Re U and Im U) recorded at a higher frequency of $f = 3.11$ MHz under the same conditions, demonstrating the frequency-independent step-like signal feature at the transition temperature.

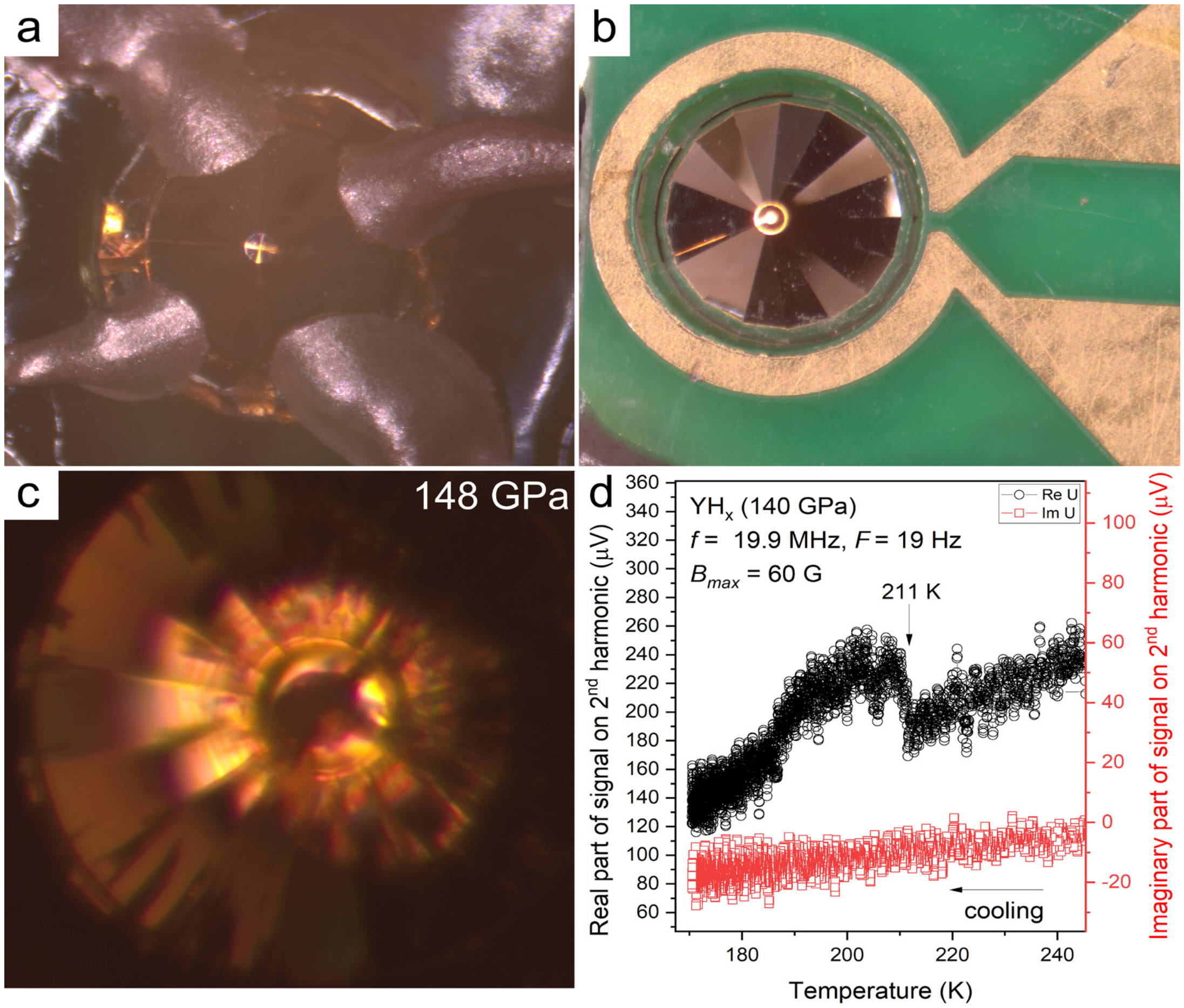

**Figure S15**. Additional RF transmission experiment for yttrium hydride $YH_x$, probably $YH_6$, at 140 ±5 GPa (DAC Y4). (a) General view of the diamond anvil cell with a 4-electrode system and Cu/Ag-based conductive adhesive used for contact bonding. (b) Lenz lens and single-turn PCB excitation coil used for the RF response detection. (c) View of the loaded sample in the chamber at high pressure. (d) Real and imaginary parts of the second harmonic (*2F*) of the modulating magnetic field ($F = 19$ Hz) when an RF signal at 19.9 MHz passes through it. Detectable $T_c \approx 211$ K in the cooling cycle.

## IV. Additional X-ray diffraction data

Synchrotron powder diffraction of the DAC Y6 sample contains a significant number of diffraction reflections and indicates the absence of higher yttrium polyhydrides. The simplest interpretation is achieved using several hexagonal structures, such as *P*6/*mmm*-$YH_{2+x}$ and *dhcp*-$YH_{2-x}$, which sometime occurs in hydrides at high pressure [36], or the recently discovered *hP*3-$YH_x$ group of hydrides [37,38]. Figures S16 and S17 show examples of such Le Bail refinements at 161 and 106 GPa. One of the phases decomposes between 121 and 106 GPa (Figure S18).

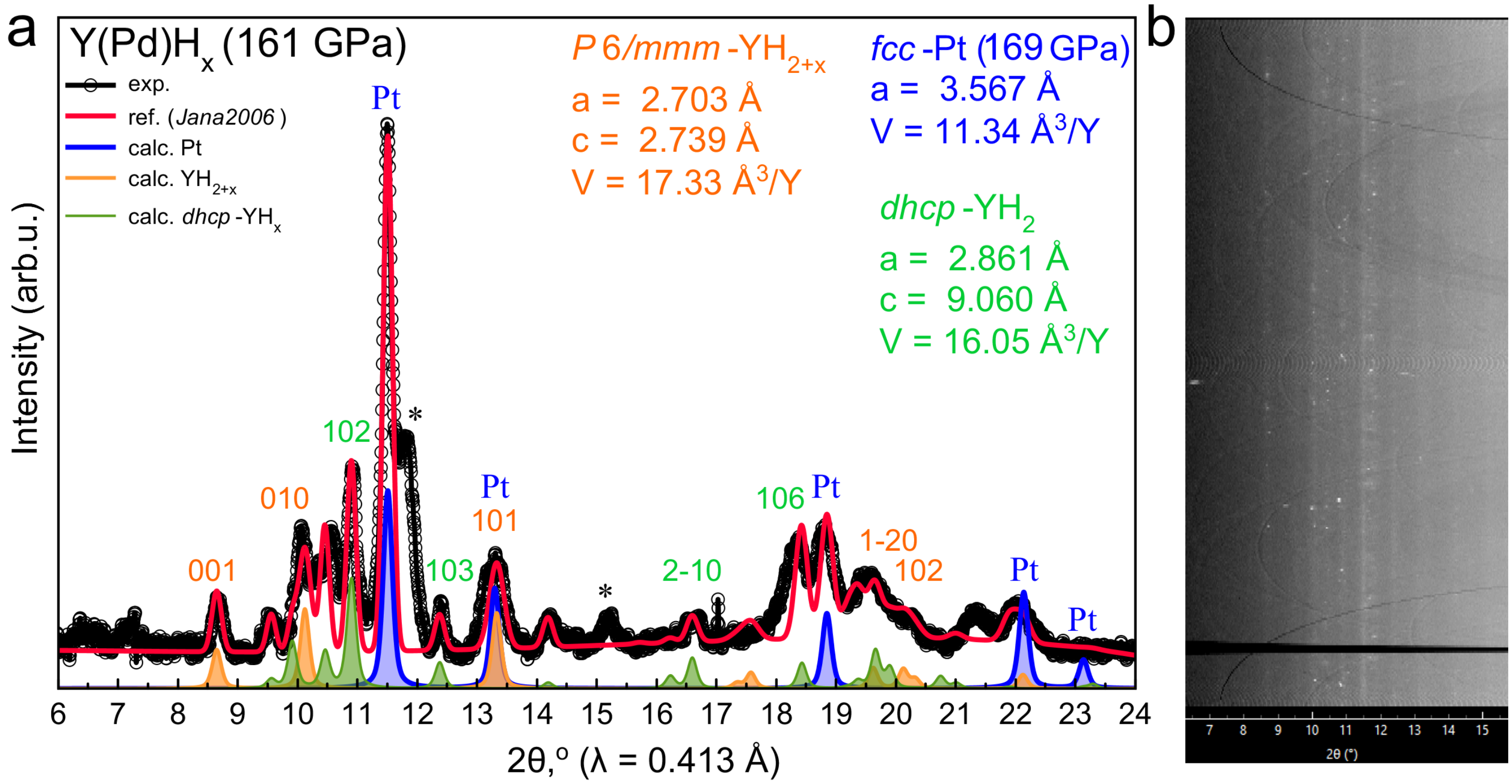

**Figure S16.** Crystal structure refinement and phase identification of Y(Pd)$H_x$ sample (DAC Y6). (a) Synchrotron X-ray diffraction pattern of the Y(Pd)$H_x$ sample measured at 161 GPa (λ = 0.413 Å). The experimental data (black open circles) are fitted with a Le Bail refinement (red curve) performed using the Jana2006 software. The individual calculated contributions of the coexisting phases are displayed at the baseline: *P*6/*mmm*-$YH_{2+x}$ phase (orange curve): a = 2.703 Å, c = 2.739 Å, V = 17.33 Å$^3$/Y. Double hexagonal close-packed *dhcp*-$YH_{2-x}$ phase (green curve): a = 2.861 Å, c = 9.060 Å, V = 16.05 Å$^3$/Y. Face-centered cubic platinum standard (*fcc*-Pt) at 169 GPa (blue curve): a = 3.567 Å, V = 11.34 Å$^3$/Y. Asterisks (*) indicate minor unindexed or impurity reflections. Miller indices (*hkl*) for the respective phases are color-coded accordingly. (b) The corresponding raw two-dimensional X-ray diffraction image recorded on the area detector as a function of the scattering angle 2θ.

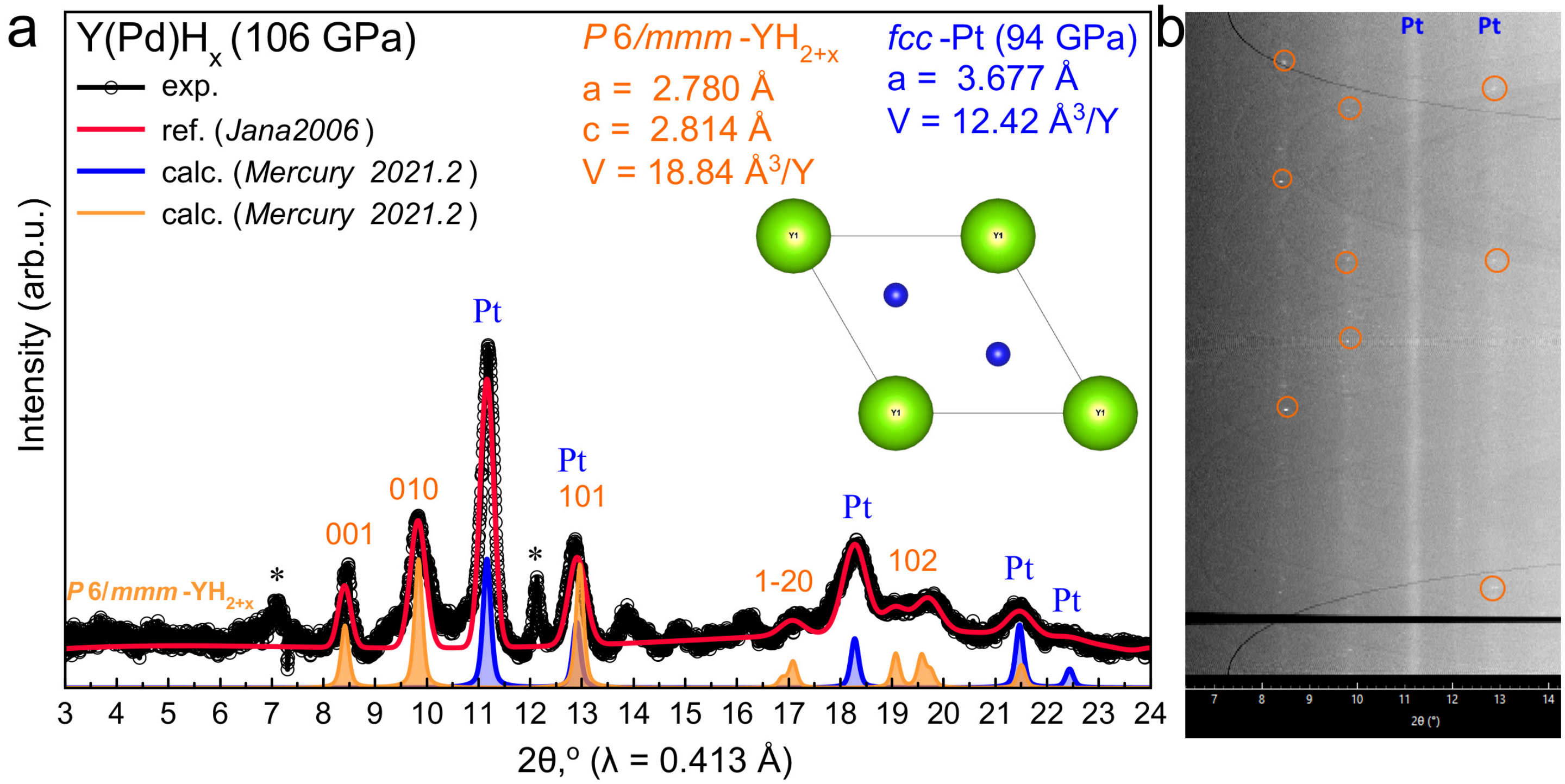

**Figure S17.** Crystal structure refinement and phase identification of Y(Pd)$H_x$ sample (DAC Y6). (a) Synchrotron X-ray diffraction pattern of the Y(Pd)$H_x$ sample measured at 106 GPa (λ = 0.413 Å). The experimental data (black open circles) are fitted with a Le Bail refinement (red curve) performed using the Jana2006 software. The individual calculated contributions of the coexisting phases are displayed at the baseline: *P*6/*mmm*-$YH_{2+x}$ phase (orange curve): a = 2.780 Å, c = 2.814 Å, V = 18.84 Å$^3$/Y. Face-centered cubic platinum standard (*fcc*-Pt, blue curve): a = 3.677 Å, V = 12.42 Å$^3$/Y. Asterisks (*) indicate minor unindexed or impurity reflections. Miller indices (*hkl*) for the respective phases are color-coded accordingly. (b) The corresponding raw two-dimensional X-ray diffraction image recorded on the area detector as a function of the scattering angle *2θ*.

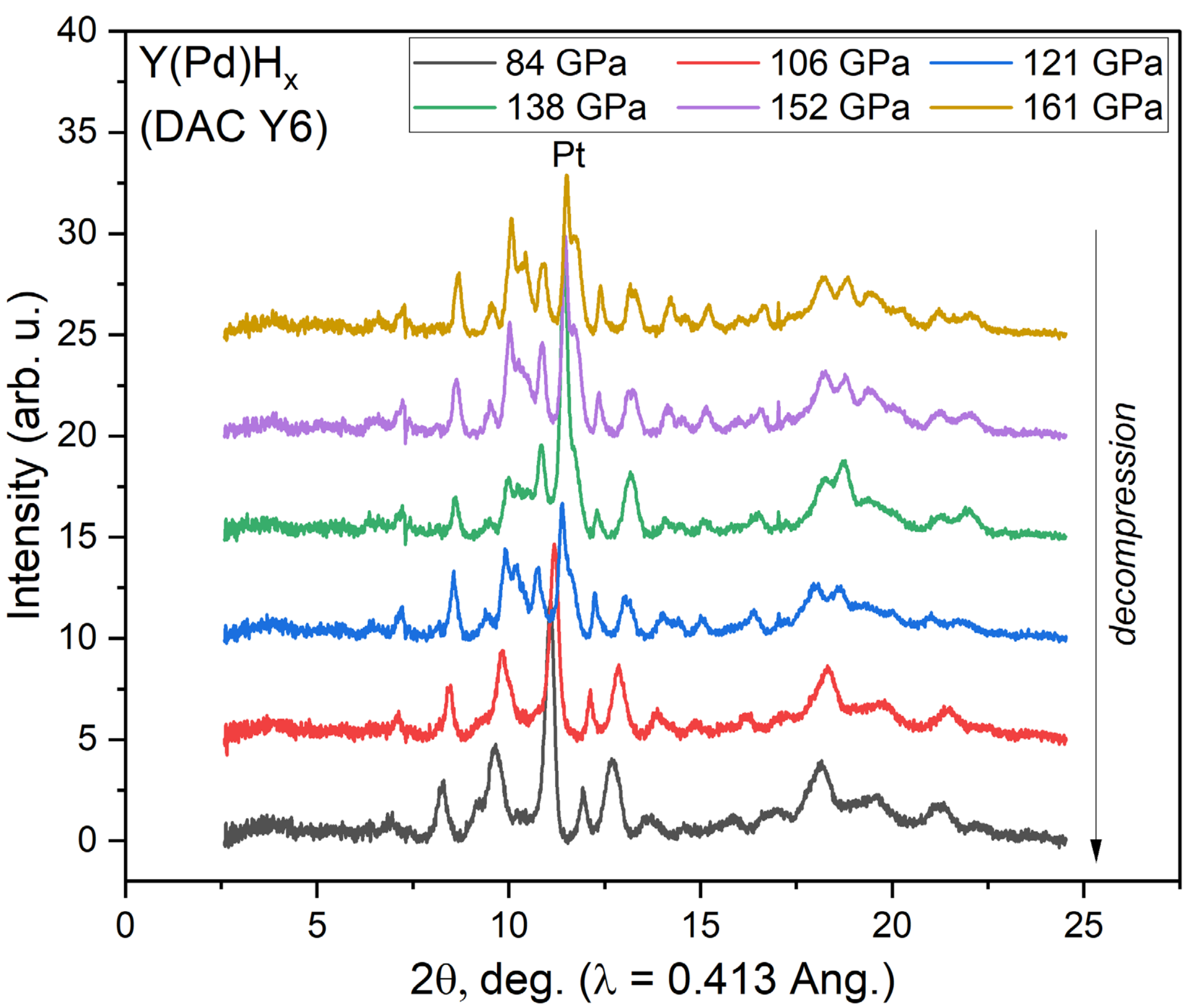

**Figure S18.** Synchrotron X-ray diffraction patterns of Y(Pd)$H_x$ sample (x = 1–3, DAC Y6) during decompression. The evolution of the XRD profiles for the Y(Pd)$H_x$ sample is shown as a function of decreasing pressure from 161 GPa down to 84 GPa. The patterns are simplified by decompression, indicating the chemical transformation (decomposition) of one of the phases (possibly, *dhcp*) in the sample as the pressure decreases.

## V. Auxiliary theoretical calculations

This section presents unconverged calculations performed using relatively sparse k- and q-grids, which were originally employed to estimate the superconducting $T_c$ during the period surrounding the discovery of $YH_9$ in 2020–2021. We used PBE (pbe-mt_fhi.UPF) pseudopotentials with k-meshes of 18×18×16 and 16×16×12, and a q-mesh density of 4×4×2. Two primary methods were applied: the gaussian broadening scheme (for $YH_9$ at 200 and 240 GPa, Figure S18a-c) and the tetrahedral integration method (for $YPdH_{12}$ and $YAlH_{12}$ at 180 GPa). In the case of ordered substitution of yttrium by aluminum or palladium in *bcc*-$Y_2H_{12}$ (Figure S19d) the tetrahedral calculations indicate that the $T_c$ in these compounds should reach substantial values of 165–210 K, comparable to that of $YH_9$. Notably, even with very coarse grids and non-ideal pseudopotentials, an acceptable order-of-magnitude estimate of the $T_c$ for $YH_9$ (~200 K) can be obtained.

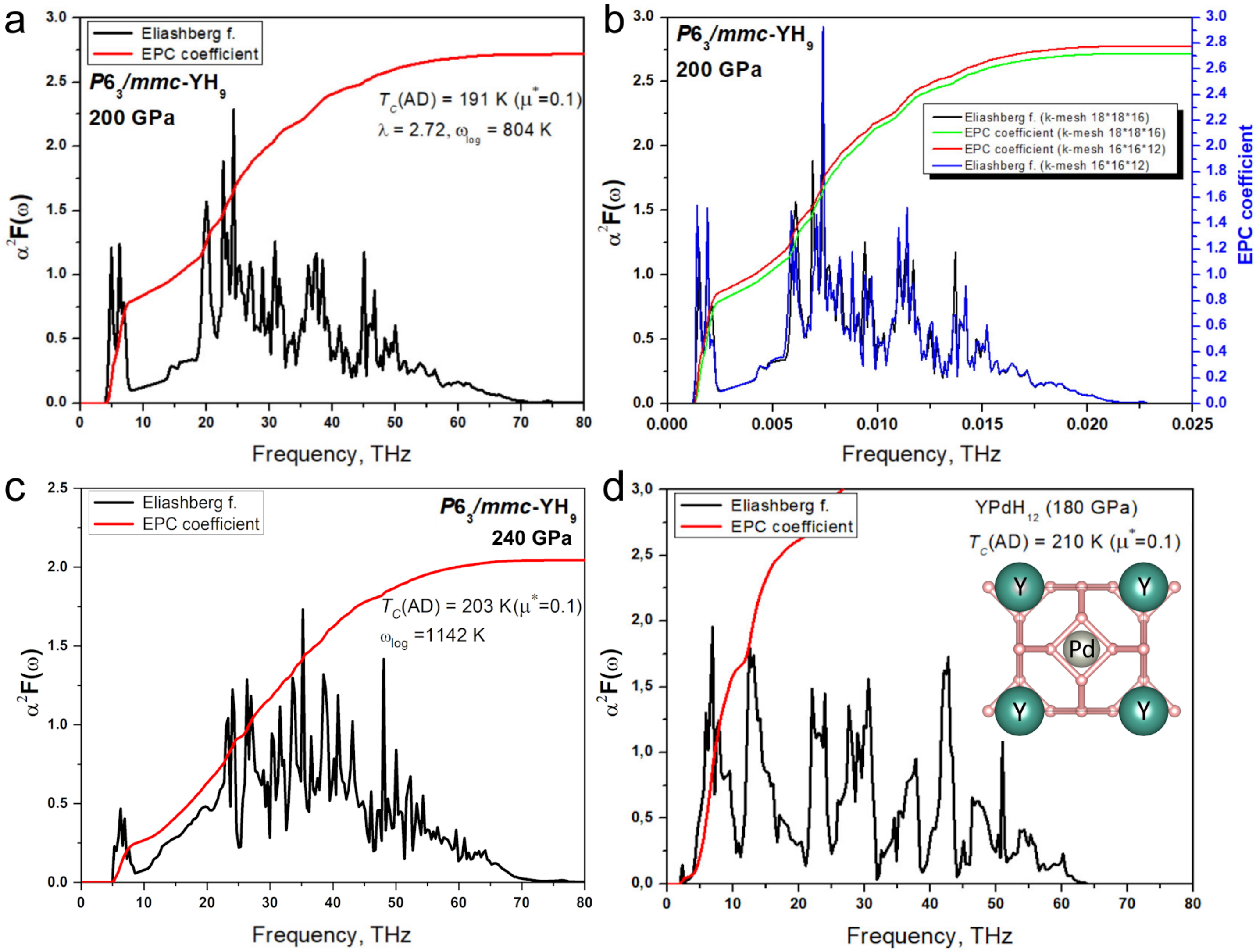

**Figure S19.** Eliashberg functions and electron-phonon interaction parameter λ for YH9 and YPdH12 at different pressures and low densities of k,q-grids. (a) YH9 at 200 GPa. (b) YH9 at 200 GPa, comparing the results of calculations with k-meshes of 18×18×16 and 16×16×12, q-mesh was 4×4×2. (c) The same calculations performed at 240 GPa. (d) YPdH12 at 180 GPa q-mesh was 2×2×2.

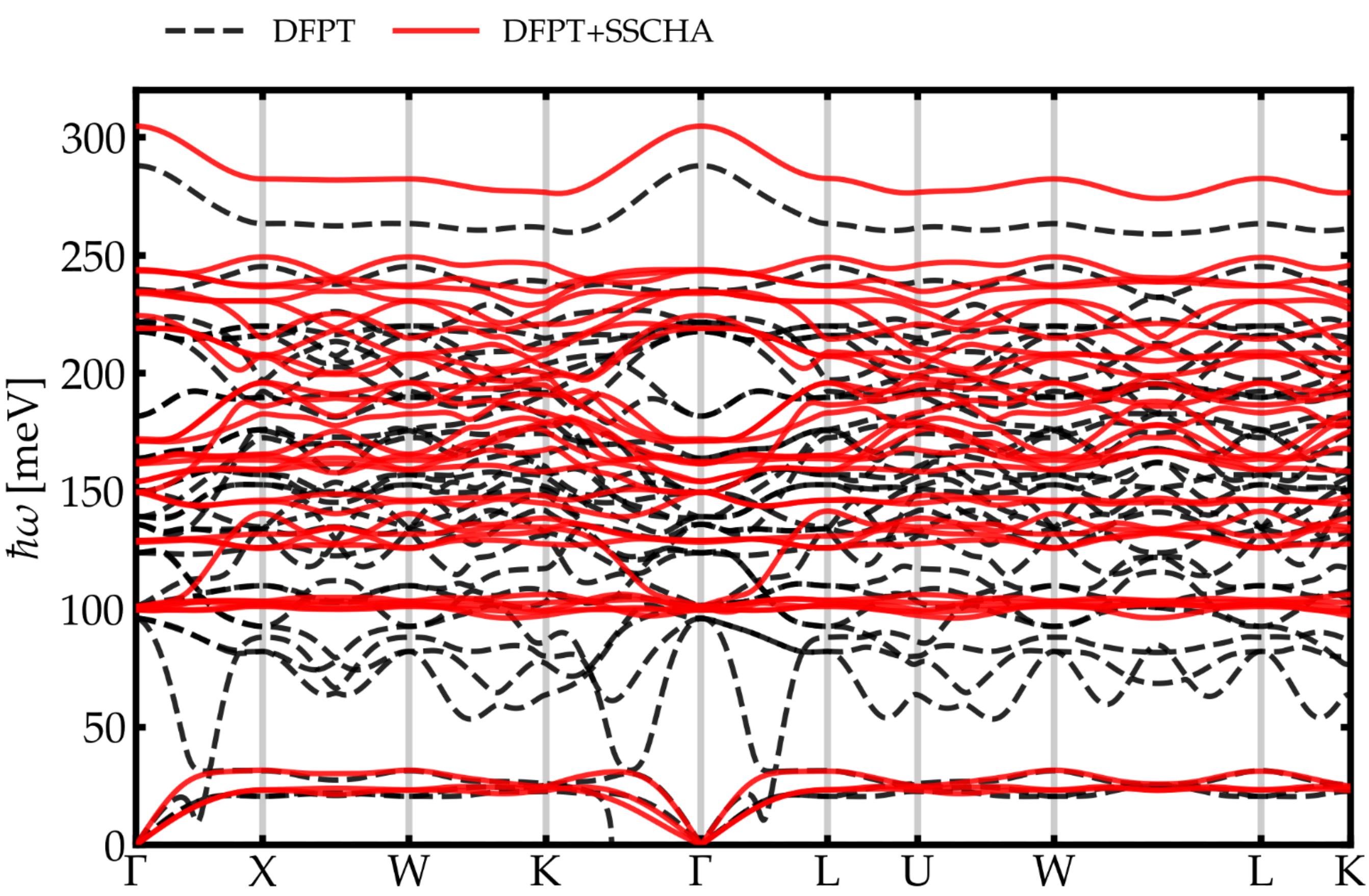

**Figure S20.** Phonon band structure of *Fm*$\overline{3}$*m*-$YH_{10}$ at 200 GPa within the harmonic approximation (black dashed lines) and including anharmonic quantum lattice effects (red solid lines).